\documentclass[useAMS,usenatbib]{mn2e}

\usepackage{graphicx}
\usepackage{subfigure}
\usepackage{paralist}
\usepackage{amssymb}
\usepackage{bm}
\usepackage{bbding}
\usepackage[fleqn]{amsmath}
\usepackage{multirow}
\usepackage{epsfig}
\usepackage{epstopdf}
\usepackage[colorlinks,
            linkcolor=blue, 
            anchorcolor=blue, 
            citecolor=blue, 
            ]{hyperref}
\title[Secular resonance]{Secular resonance of inner test particles in hierarchical planetary systems}
\author[Lei]{Hanlun Lei$^{1,2}$\thanks{E-mail: leihl@nju.edu.cn}\\
$^{1}$ School of Astronomy and Space Science, Nanjing University, Nanjing 210023, China\\
$^{2}$ Key Laboratory of Modern Astronomy and Astrophysics in Ministry of Education, Nanjing University, Nanjing 210023, China}

\begin{document}

\date{Accepted. Received; in original form}

\pagerange{\pageref{firstpage}--\pageref{lastpage}} \pubyear{2020}

\maketitle
\label{firstpage}

\renewcommand{\arraystretch}{1.3}

\begin{abstract}
The present work studies the secular resonance associated with the critical argument $\sigma = \varpi$ ($\varpi$ is the longitude of pericentre) for inner test particles moving in low-eccentricity region with inclination $i$ smaller than $39^{\circ}$. To formulate the dynamical model, the double-averaged Hamiltonian is formulated up to an arbitrary order in the semimajor axis ratio, and then those high-order periodic terms are removed from the double-averaged Hamiltonian by means of Hori--Deprit transformation technique. The resulting Hamiltonian determines a resonant model with a single degree of freedom. Based on the resonant model, it becomes possible to explore the phase-space structure, resonant centre, and resonant width in an analytical manner. In particular, an excellent correspondence is found between the resonant width in terms of the eccentricity variation and the maximum variation of eccentricity ($\Delta e$) for test particles initially placed on quasi-circular orbits. It means that the secular dynamics in the low-eccentricity space with $i < 39^{\circ}$ is dominantly governed by the secular resonance associated with $\sigma = \varpi$.
\end{abstract}


\section{Introduction}
\label{Sect1}

The Hamiltonian of a hierarchical planetary system is composed of two Keplerian terms plus a coupling term (also called the disturbing function) describing the mutual interaction between the inner and outer planets \citep{harrington1969stellar}. To study the secular dynamics on timescales much longer than the orbital periods, it is usual to filter out the short-period effects by averaging the planetary disturbing function over the orbital periods of the inner and outer planets. After averaging, the Keplerian energies of the inner and outer planets are separately conserved, and the secular evolution of system is a dynamical outcome of their angular momentum exchange \citep{naoz2016eccentric}. The averaging approximation can be realised by means of Von Zeipel transformation \citep{naoz2013secular}.

The conventional double averaging corresponds to the secular approximation at the first order in the magnitude of perturbation, indicating that the approximation works well for those systems with high hierarchy (i.e., the systems with weak perturbation). Physical hierarchy is measured by the separability among degrees of freedom of the dynamical system or the differences between the periods of the so-called ``short-period'' and ``long-period'' variables \citep{beauge2006high}, so that high hierarchy requires that the semimajor axis ratio is small. However, if a planetary three-body system is not highly hierarchical (in this situation the perturbation is so strong that the separation of `short period' and `long period' becomes blurred), the short-period effects within the orbital periods of the outer and/or inner planets need to be taken into account in the secular approximation \citep{cuk2004secular, luo2016double, lei2018modified, lei2019semi, hamers2019analytica, hamers2019analyticb} or high-order secular Hamiltonian needs to be formulated to predict long-term behaviours \citep{beauge2006high}.

As for highly hierarchical planetary systems, the conventional double-averaging approach has been widely used to investigate the secular dynamics. Considering one of the inner objects as a test particle, both \citet{kozai1962secular} and \citet{lidov1962evolution} expanded the disturbing function as a power series of the semimajor axis ratio ($\alpha$) between the test particle and the perturber and considered applications when the orbit of the outer binary is circular. \citet{kozai1962secular} investigated the secular influences of Jupiter's perturbations upon the main belt asteroids in our Solar system, and \citet{lidov1962evolution} considered the secular evolution of artificial satellites around the Earth perturbed by the Sun and Moon. In both the landmark works, the double-averaged disturbing function (or Hamiltonian) is independent on the longitude of ascending node due to the axisymmetric potential, making the $z$-axis component of the angular momentum be conserved (here the $z$-axis is along the vector of the total angular momentum) and the resulting dynamical model be integrable. Under the quadrupole-level approximation, a well-known secular resonance (also called Kozai resonance) occurs between the longitude of pericentre and the longitude of ascending node when the mutual inclination is higher than $39.2^{\circ}$ \citep{kozai1962secular, thomas1996kozai}. Due to the Kozai resonance, the eccentricity and inclination undergo a large coupled oscillation, and such a mechanism of exciting eccentricity and inclination is called the Kozai--Lidov (KL) mechanism. In recent years, many researchers relaxed the assumptions made in \citet{kozai1962secular} and \citet{lidov1962evolution} in terms of the following three aspects: (a) the outer orbit is eccentric, (b) all three objects involved are not massless, and (c) the disturbing function is relaxed to the octupole (or higher) orders in the semimajor axis ratio \citep{michtchenko2006modeling, libert2007exoplanetary, libert2009kozai, migaszewski2009equilibria, naoz2011hot, naoz2013secular, volpi20193d}. Under these relaxations, the conventional KL mechanism is extended to the eccentric Kozai--Lidov (EKL) mechanism. Both the KL and EKL mechanisms have been widely utilised to interpret secular phenomena in practical astrophysical systems \citep{soderhjelm1982studies, thomas1996kozai, holman1997chaotic, blaes2002kozai, miller2002four, wu2003planet, michtchenko2006modeling, libert2007exoplanetary, libert2009kozai, migaszewski2009equilibria, lithwick2011eccentric, katz2011long, naoz2011hot, naoz2013secular, shappee2013mass, antognini2014rapid, antognini2015timescales, petrovich2015hot, naoz2016eccentric, carvalho2016exoplanets, naoz2017eccentric, sidorenko2018eccentric, volpi20193d}.

Based on the aforementioned discussions, we know that, in a hierarchical planetary system with the disturbing body on an eccentric orbit, the secular dynamics at mutual inclinations higher than $\sim$$39.2^{\circ}$ is dominantly governed by Kozai resonance. A question arises: what's the dynamical mechanism in governing the secular dynamics in the parameter space where Kozai resonance cannot happen (i.e., in the region with inclinations smaller than $\sim$$39^{\circ}$)? To my knowledge, there are few publications focusing on this topic.

\citet{funk2011influence} made a parametric study in the region with $i < 39^{\circ}$ for the long-term stability of inclined and quasi-circular test particles (fictitious Earth-like planets) moving under the gravity field produced by a central star and an eccentric giant planet. They found an interesting dynamical region around the mutual inclination of $\sim$$35^{\circ}$ where the low-eccentricity orbits are long-time stable, indicating that the inclined ($\sim$$35^{\circ}$) quasi-circular Earth-mass companion is possible to survive in the habitable zone of extrasolar systems \citep{funk2011influence}. Under the double-averaged model at the octupole-level approximation, \citet{libert2012interesting} computed the maximum variation of eccentricity as a function of the mutual inclination (see Fig. 6 in their work) and reproduced the long-time stable region at inclination of $\sim$$35^{\circ}$. To understand the long-term stable dynamics in this special region, \citet{libert2012interesting} adopted the 12th-order (double-averaged) Hamiltonian expansion in eccentricities and inclinations \citep{libert2007exoplanetary} and formulated a second-order secular Hamiltonian by means of Lie-series transformation method to determine the fundamental frequencies, similar to the method used in \citet{libert2008secular}. Then, \citet{libert2012interesting} showed that this region presents an equality of two fundamental frequencies (please refer to Fig. 8 in their work), and thus they concluded that the long-term stable dynamics at inclination of $\sim$$35^{\circ}$ is due to the existence of the secular resonance associated with $\sigma = \omega - \Omega$.

\begin{figure}
\centering
\includegraphics[width=0.48\textwidth]{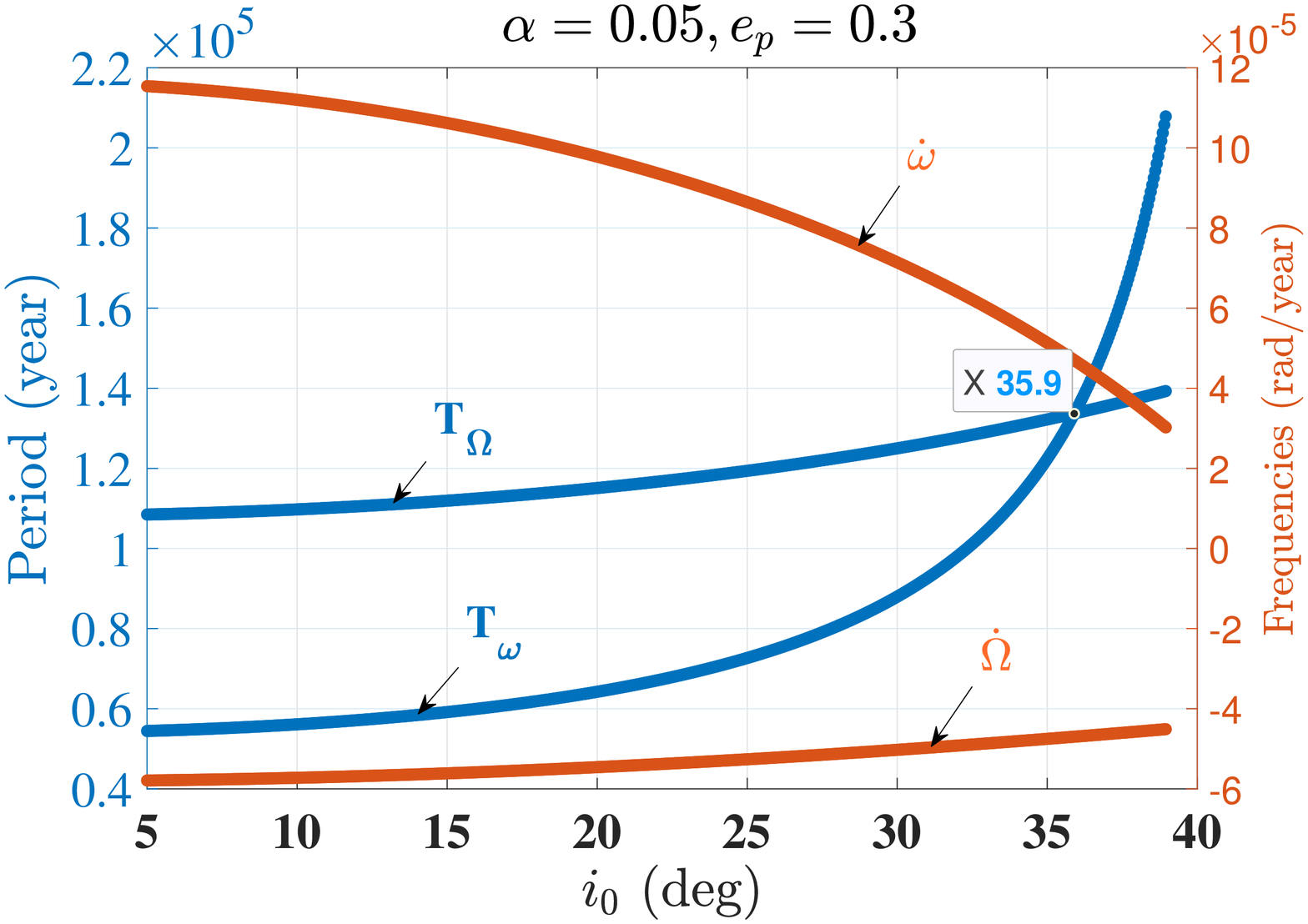}
\caption{Fundamental frequencies ($\dot \omega$ and $\dot \Omega$) as well as their respective periods ($T_{\omega}$ and $T_{\Omega}$) as functions of initial inclination $i_0$. In simulations, the double-averaged Hamiltonian shown by equation (\ref{Eq2}) is truncated at the sixth order in the semimajor axis ratio $\alpha$ and the secular Hamiltonian is formulated up to the second order in $\varepsilon$ by means of Lie-series transformation (see Section \ref{Sect4} for the definition of parameter $\varepsilon$ and the Lie-series transformation). The planetary system consists of a central body with mass $m_{\rm Sun}$, a giant planet with mass $m_{\rm Jupiter}$ and a test particle with infinitesimal mass. The same parameters as the ones in \citet{libert2012interesting} are used in the simulations: the eccentricity of the giant planet is assumed as $e_p = 0.3$, the semimajor axis ratio as $\alpha = 0.05$ and the initial eccentricity of the test particle as $e = 1 \times 10^{-6}$.}
\label{Fig1}
\end{figure}

Based on our double-averaged Hamiltonian in power series of the semimajor axis ratio (to be presented in Section \ref{Sect2}), we formulated the secular Hamiltonian by using the same transformation method adopted by \citet{libert2012interesting} and reproduced the fundamental frequencies ($\dot \omega$ and $\dot \Omega$) as well as the corresponding periods ($T_{\omega}$ and $T_{\Omega}$) for test particles moving in low-eccentricity regions. The fundamental frequencies as well as their respective periods are reported in Fig. \ref{Fig1}, where the secular Hamiltonian is formulated up to the second order in $\varepsilon$ (see Section \ref{Sect4} for the definition of parameter $\varepsilon$). It is observed that the equality between $T_{\Omega}$ and $T_{\omega}$ occurs at inclination of $\sim$$35.9^{\circ}$. The curves of $T_{\omega}$ and $T_{\Omega}$ given in Fig. \ref{Fig1} are in good agreement with the ones shown in \citet{libert2012interesting} (see Fig. 8 in their work). However, in our simulations the fundamental frequencies holds $\dot \omega > 0$ and $\dot \Omega < 0$, so that the intersection point between the curves of $T_{\Omega}$ and $T_{\omega}$ may lead us to a different conclusion: the secular resonance associated with $\sigma = \omega + \Omega = \varpi$ (rather than $\sigma = \omega - \Omega$) occurs at the inclination of the intersection point.

The purpose of this work is to understand how the secular resonance dominates the secular dynamics in the parameter space with inclinations smaller than $39^{\circ}$ by studying the secular resonance with critical argument $\sigma = \omega + \Omega$. To this end, it is required to formulate the resonant Hamiltonian and then explore phase-space structures, resonant centres and widths. In the process of formulating resonant Hamiltonian, we adopt the Hori--Deprit transformation technique \citep{hori1966theory, deprit1969canonical} to remove high-order short-period effects from the Hamiltonian. Based on the resulting resonant Hamiltonian, the phase-space structures, resonant centres and widths of the resonances are analyzed. Interestingly, there is an excellent correspondence between the resonant width in terms of the eccentricity variation and the maximum variation of eccentricity ($\Delta e$) for particles initially placed on quasi-circular orbits. It means that the shape of $\Delta e$ numerically determined is sculpted by the secular resonance associated with $\sigma = \omega + \Omega$.

The structure of the present work is organised as follows. In Section \ref{Sect2}, the double-averaged Hamiltonian is briefly introduced and, in Section \ref{Sect3}, secular dynamics in the low-eccentricity region with inclinations smaller than $39^{\circ}$ is numerically explored. In Section \ref{Sect4}, the perturbation theory based on the Hori--Deprit transformation technique is introduced and the second-order resonant Hamiltonian is formulated. Results and discussions are provided in Section \ref{Sect5}. Finally, Section \ref{Sect6} concludes this work.

\section{Hamiltonian function}
\label{Sect2}

\begin{figure*}
\centering
\includegraphics[width=0.4\textwidth]{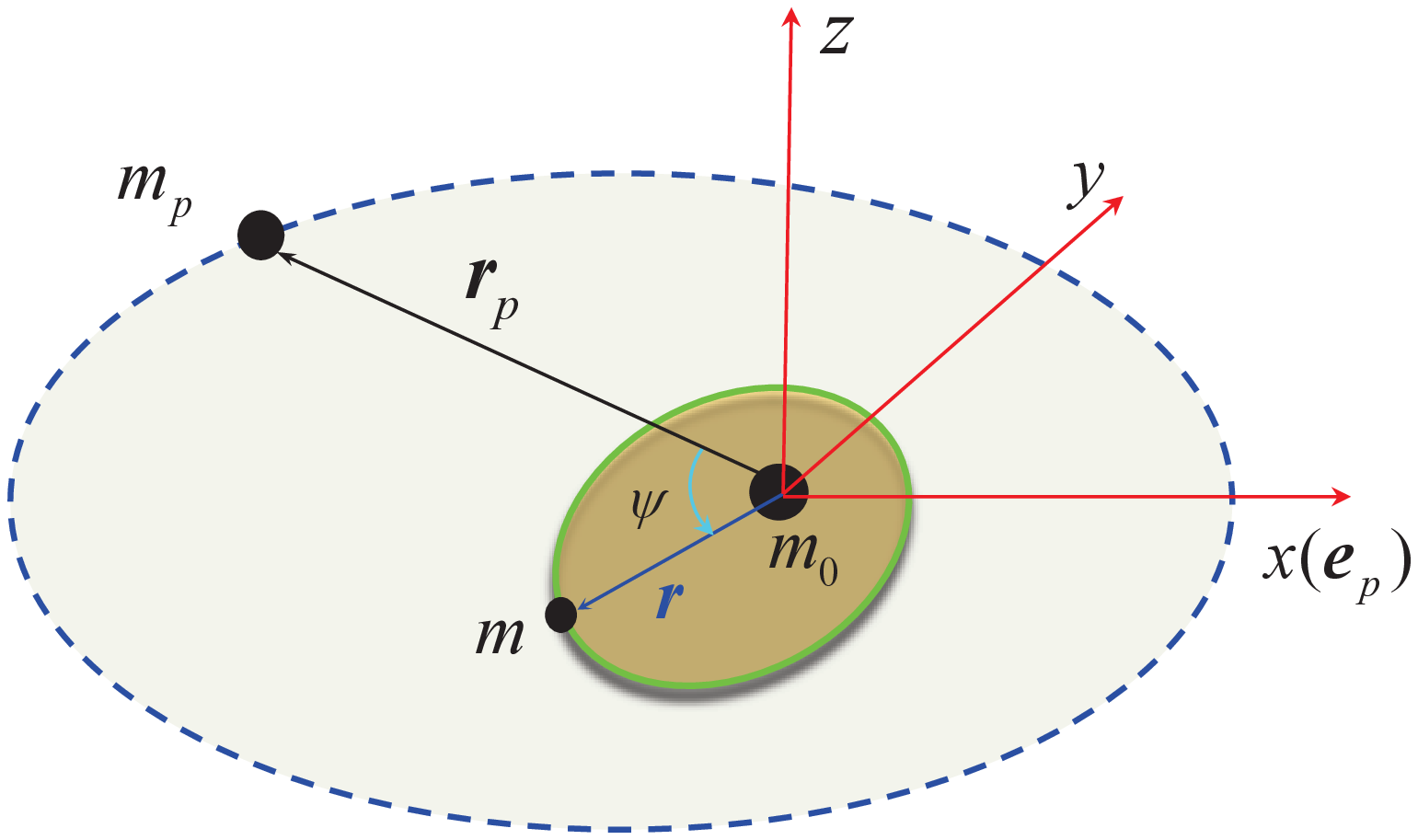}
\includegraphics[width=0.45\textwidth]{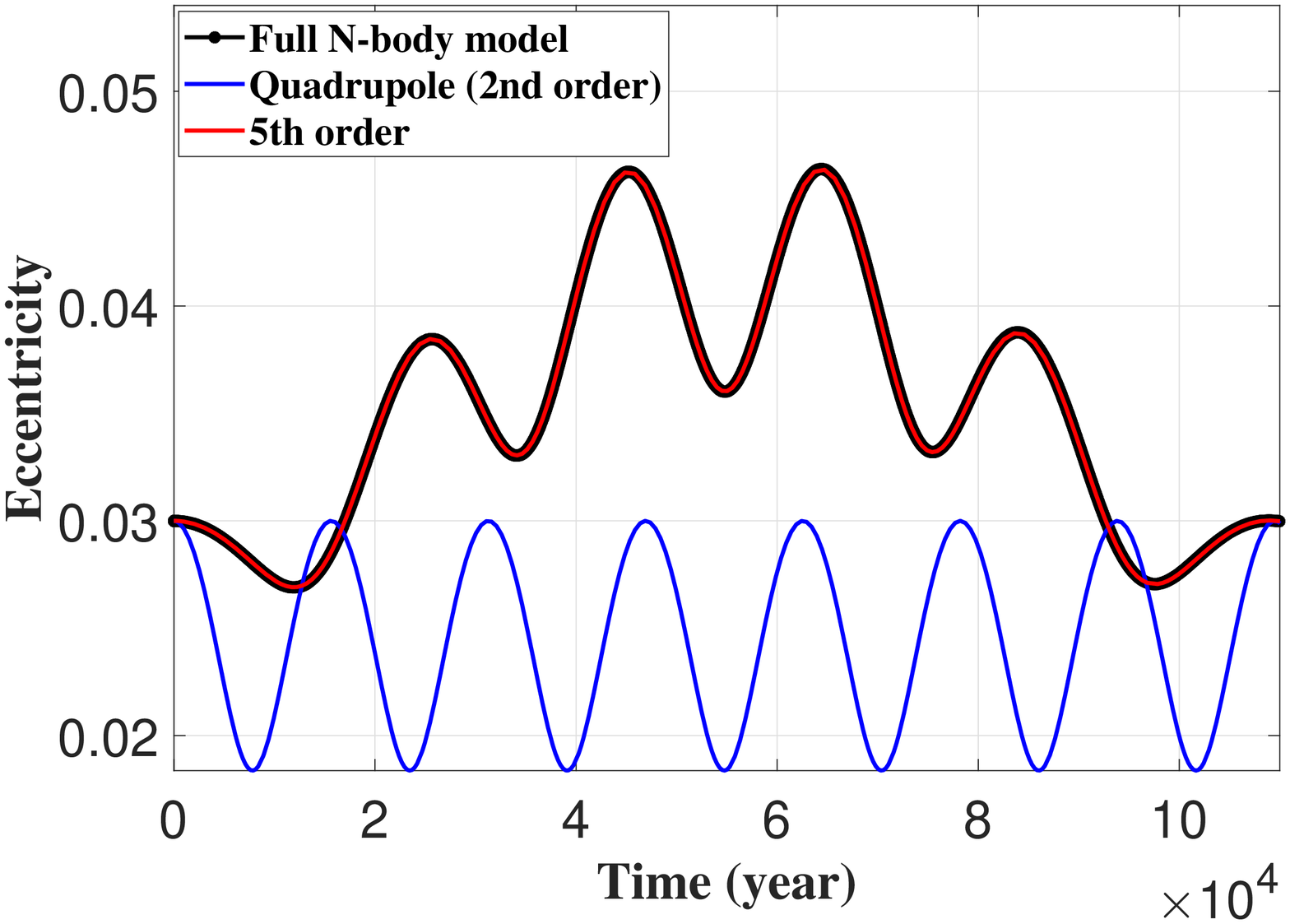}
\caption{Relative configuration of three objects in a hierarchical planetary system and the definition of coordinate system adopted in this study (\emph{left panel}) and time histories of the eccentricity for numerically integrated trajectories (\emph{right panel}). In the left panel, the position vectors of the test particle and the disturbing body are $\bm{r}$ and ${\bm r}_p$. In the right panel, the trajectories are numerically propagated under the full $N$-body model (i.e. the elliptic restricted three-body problem) as well as under the double-averaged models truncated at different orders in the semimajor axis ratio between the inner test particle and giant planet $\alpha = a/a_p$. The dynamical model is characterised by $m_0 = m_{\rm{Sun}}$, $a_p = 1.0$ $\rm{au}$, $m_p = m_{\rm{Jupiter}}$ and $e_p = 0.3$. The initial elements are taken as follows: $a_0=0.1$ $\rm{au}$, $e_0=0.03$, $i_0=30^{\circ}$, $\Omega_0 = 90^{\circ}$ and $\omega_0 = 270^{\circ}$.}
\label{Fig2}
\end{figure*}

The hierarchical planetary system considered in this study is composed of a central body with mass $m_0$ (e.g., a star), a terrestrial planet with mass $m$ (e.g., an Earth-mass planet) and a faraway disturbing body with mass $m_p$ (e.g., a Jupiter-mass planet). In this system, the Earth-mass planet and the central body constitute a tight inner binary, and the disturbing body and the barycentre of the inner binary form an outer binary (please refer to the left panel of Fig. \ref{Fig2} for the relative configuration). Due to the fact that $m$ is much smaller than $m_p$ and $m_0$, the Earth-mass planet in this planetary system can be approximated as a test particle. In the test particle approximation, the hierarchical planetary system reduces to an elliptic restricted three-body problem, where the faraway disturbing body moves around the central body on a fixed elliptic orbit.

The hierarchy of the planetary system is specified by the semimajor axis ratio between the inner and outer planets, given by $\alpha = a/a_p$. For small $\alpha$, the Hamiltonian function, describing the mutual interaction between the orbits of the inner and outer planets, can be expanded as a power series of $\alpha$. As a result, the complete Hamiltonian that describes the motion of test particles can be written as \citep{morbidelli2002modern}
\begin{equation}\label{Eq1}
{\cal H} =  - \frac{\mu }{{2a}} - \frac{{{\cal G}{m_p}}}{{{a_p}}}\sum\limits_{n = 2}^\infty  {{\alpha ^n}{{\left( {\frac{r}{a}} \right)}^n}{{\left( {\frac{{{a_p}}}{{{r_p}}}} \right)}^{n + 1}}{P_n}\left( {\cos \psi } \right)},
\end{equation}
where $\mu = {\cal G} {m_0}$ with $\cal{G}$ as the universal gravitation constant, $r$ and $r_p$ are, respectively, the magnitudes of the position vectors $\bm{r}$ for the test particle and $\bm{r}_p$ for the disturbing body, $\psi$ is the relative angle between $\bm{r}$ and $\bm{r}_p$, and ${P_n}\left( {\cos \psi } \right)$ are the Legendre polynomial of $\cos \psi$ with degree $n$. Unless otherwise stated, in the entire work we use the variables with subscript `$p$' for the wide orbit of the perturber and the ones without any subscript for the orbit of test particles.

To describe the motion of the objects involved in the planetary system, we choose an inertial right-handed reference frame originated at the central body, denoted by $o$-$xyz$, with the invariable plane of the system (i.e. the orbit of the disturbing body around the central body) as the fundamental plane, the total angular momentum vector as the direction of the $z$-axis and the eccentricity vector of the disturbing body as the direction of the $x$-axis (please refer to the left panel of Fig. \ref{Fig2} for details). Under the coordinate system, the orbits of the test particle (perturber) are described by the semimajor axis $a$ ($a_p$), eccentricity $e$ ($e_p$), inclination $i$ ($i_p$), longitude of ascending node $\Omega$ ($\Omega_p$), the argument of pericentre $\omega$ ($\omega_p$) and mean anomaly $M$ ($M_p$). The longitude of pericentre is defined by $\varpi = \omega + \Omega$ for the test particle and $\varpi_p = \omega_p + \Omega_p$ for the perturber. Due to the specific choice of the fundamental plane about the coordinate system, it is not difficult to get $i_p = 0$.

The Hamiltonian function given by equation (\ref{Eq1}) can be expanded as a Fourier series form \citep{kaula1961analysis, kaula1962development, murray1999solar, beauge2006high, lei2019semi}. To study the secular evolution, it is usual to perform double averages for the Hamiltonian function over the orbital periods of the inner and outer planets \citep{libert2009kozai, laskar2010explicit, lithwick2011eccentric,  naoz2011hot, naoz2013secular, li2014chaos, carvalho2016exoplanets, naoz2017eccentric, lei2018modified, lei2019semi, lei2020dynamical}. The phase averaging, also called the secular approximation \citep{naoz2016eccentric}, is realised by Von Zeipel transformation, as discussed in the Appendix of \citet{naoz2013secular}. Eliminating those terms involving the mean anomalies $M$ or $M_p$ from the Fourier series expression of Hamiltonian function \citep{lei2019semi}, the double-averaged formulation of Hamiltonian can be written as (for simplicity, we still use ${\cal H}$ to represent the double-averaged Hamiltonian)
\begin{equation}\label{Eq2}
\begin{aligned}
{\cal H}  =&  - \frac{\mu }{{2a}} - \frac{{{\cal G}{m_p}}}{{{a_p}}}\sum\limits_{n = 2}^N {\sum\limits_{k = 0}^{\left[ {\frac{n}{2}} \right]} {\sum\limits_{q = 0}^{n - 2k} {\sum\limits_{{t_1} = 0}^{n - 2k - q} {\sum\limits_{{t_2} = 0}^q {\sum\limits_{{t_3} = 0}^{n - 2k - q} {\sum\limits_{{t_4} = 0}^q {\kappa_1 \kappa_2 } } } } } } } \\
&\times {\alpha ^n} X_0^{n,\left( {n - 2k - 2{t_1} - 2{t_2}} \right)}\left( e \right)X_0^{ - \left( {n + 1} \right),\left( {n - 2k - 2{t_3} - 2{t_4}} \right)}\left( {{e_p}} \right)\\
&\times {\cos ^q} i \cos{\theta},
\end{aligned}
\end{equation}
where $\left[ {\frac{n}{2}} \right]$ represents the lower integer of $\frac{n}{2}$, $X_0^{a,b}(e)$ are the Hansen coefficients, and the angle $\theta$ is given by
\begin{equation*}
\begin{aligned}
\theta &= \left( {n - 2k - 2{t_1} - 2{t_2}} \right)\varpi + \left({n - 2k - 2{t_3} - 2{t_4}} \right) {\varpi}_p\\
&- 2\left( {n - 2k - {t_1} - {t_2} - {t_3} - {t_4}} \right)\Omega,
\end{aligned}
\end{equation*}
where the d'Alembert rule holds. Due to the specific choice of the $x$-axis about the coordinate system (along the eccentricity vector of the perturber), it is not difficult to get $\varpi_p = 0$. Consequently, the angle $\theta$ becomes
\begin{equation*}
\theta = \left( {n - 2k - 2{t_1} - 2{t_2}} \right) \varpi - 2\left( {n - 2k - {t_1} - {t_2} - {t_3} - {t_4}} \right)\Omega.
\end{equation*}
In addition, the coefficients $\kappa_1$ and $\kappa_2$ in equation (\ref{Eq2}) are given by
\begin{equation*}
{\kappa _1} = \frac{{{{\left( { - 1} \right)}^{k + 3q + {t_2} + {t_4}}}}}{{{2^{3n - 4k}}}} \frac{{(2n - 2k)!}}{{k!(n - k)!(n - 2k)!}},
\end{equation*}
and
\begin{equation*}
{\kappa _2} ={n-2k \choose q} {n - 2k - q \choose {t_1}}{q \choose {t_2}}{n - 2k - q \choose {t_3}}{q \choose {t_4}}
\end{equation*}
with the binomial coefficients defined by ${n \choose m} = {\frac{{n!}}{{m!(n - m)!}}}$.

When the truncated order is set as $N=2$, the Hamiltonian ${\cal H}$ given by equation (\ref{Eq2}) stands for the quadrupole-level approximation and, when the truncated order is assumed at $N=3$, the associated Hamiltonian ${\cal H}$ corresponds to the octupole-level approximation, and the like.

Some remarks are made here for the expansion of disturbing function. About the coplanar and non-resonant two-planet systems, \citet{lee2003secular} expanded the double-averaged planetary disturbing function up to the third order in terms of the ratio of semimajor axes to develop the octupole-level secular perturbation theory. The octupole-level approximation of disturbing function can be found in \citet{libert2012interesting}. As a generalisation of the octupole-level secular theory, \citet{migaszewski2008secular} formulated the double-averaged planetary disturbing function up to the 24th order with respect to the semimajor axis ratio. Extending to the spatial two-planet systems, \citet{laskar2010explicit} revisited the derivation of the double-averaged planetary disturbing function up to an arbitrary order in terms of the ratio of the semi-major axes. In particular, when one of the two planets is massless (it reduces to a restricted three-body problem), we believe that the expansion given by \citet{laskar2010explicit} should be equivalent to the one shown by equation (\ref{Eq2}).

As usual, we adopt a set of canonical variables, known as modified Delaunay's elements, to describe the motion of test particles \citep{morbidelli2002modern},
\begin{equation*}
\begin{aligned}
&\Lambda = \sqrt{\mu a}, \quad \lambda = M + \varpi,\\
&P = \sqrt{\mu a}\left(\sqrt{1-e^2}-1\right),\quad p = \varpi,\\
&Q = \sqrt{\mu a (1-e^2)} \left(\cos{i} - 1\right),\quad q = \Omega.
\end{aligned}
\end{equation*}
It is noted that the variables $p$, $q$, $P$ and $Q$ defined here have opposite signs in comparison to the traditional ones (please see \citet{morbidelli2002modern} for the classical version). In the double-averaged environment, the angular variable $\lambda$ disappears from the Hamiltonian, so that the Keplerian energy (or the semimajor axis) of the test particle remains stationary in the long-term evolution. Thus, the first term in the Hamiltonian given in equation (\ref{Eq2}) can be removed.

For the sake of simplicity, we organise the double-averaged Hamiltonian in a compact form:
\begin{equation}\label{Eq4}
{\cal H} = \sum\limits_{{k_1} = 0}^N {\sum\limits_{{k_2 (2)} =  - 2N}^{2N} {{{\cal C}_{{k_1},{k_2}}}\cos \left( {{k_1}p + {k_2}q} \right)}},
\end{equation}
where $k_2 (2)$ means the index $k_2$ changes from $-2N$ to $2N$ with step 2 (i.e., $k_2$ should be an even number) and the coefficients ${{{\cal C}_{{k_1},{k_2}}}}$ are functions of the action variables ($P$ and $Q$) and their expressions can be directly derived from equation (\ref{Eq2}).

Replacing the double-averaged Hamiltonian in the canonical relations, we can obtain the secular equations of motion as follows:
\begin{equation}\label{Eq5}
\begin{aligned}
\dot p = & \frac{{\partial {\cal H}}}{{\partial P}},\quad \dot P =  - \frac{{\partial {\cal H}}}{{\partial p}},\\
\dot q = & \frac{{\partial {\cal H}}}{{\partial Q}},\quad \dot Q =  - \frac{{\partial {\cal H}}}{{\partial q}},
\end{aligned}
\end{equation}
which determines a dynamical system with two degrees of freedom. In this work, we call the model shown by equation (\ref{Eq5}) the full secular model.

To validate the full secular model truncated at different order in $\alpha$, we introduce the full $N$-body model (i.e., the elliptic restricted three-body problem), in which the equations of motion for test particles are given by
\begin{equation}\label{Eq5-1}
\ddot {\bm r} =  - \mu \frac{\bm r}{{{r^3}}} - {\cal G}{m_p}\left( {\frac{{{\bm r} - {{\bm r}_p}}}{{{{\left| {{\bm r} - {{\bm r}_p}} \right|}^3}}} + \frac{{{{\bm r}_p}}}{{r_p^3}}} \right).
\end{equation}
The Hamiltonian of the full $N$-body model is provided by equation (\ref{Eq1}).

Integrating the differential equations given by equations (\ref{Eq5}) and (\ref{Eq5-1}) numerically, we can produce the trajectories in the full secular model and in the full $N$-body model. Figure \ref{Fig2} shows time histories of the eccentricities of the trajectories numerically propagated under the full secular models truncated at different orders in $\alpha$ as well as under the full $N$-body model. Please refer to the caption for more details about the setting of initial conditions. The parameters specifying the dynamical model are $m_0 = m_{\rm Sun}$, $m_p = m_{\rm Jupiter}$, $a_p = 1$ $\rm au$ and $e_p = 0.3$ (unless otherwise specified, in the entire work we adopt this model to perform practical simulations).

From Fig. \ref{Fig2}, it is observed that (a) there is an obvious deviation for the evolution of eccentricity under the quadrupole-order approximation compared to that produced under the full $N$-body model, meaning that the quadrupole-level approximation (i.e., 2nd order in $\alpha$) cannot predict the long-term behaviours in the full $N$-body model and (b) the curve of eccentricity produced under the full secular model truncated at the 5th order in $\alpha$ is in perfect agreement with that produced under the full $N$-body model.

\section{Secular dynamics at inclinations below $39^{\circ}$}
\label{Sect3}

\begin{figure*}
\centering
\includegraphics[width=0.33\textwidth]{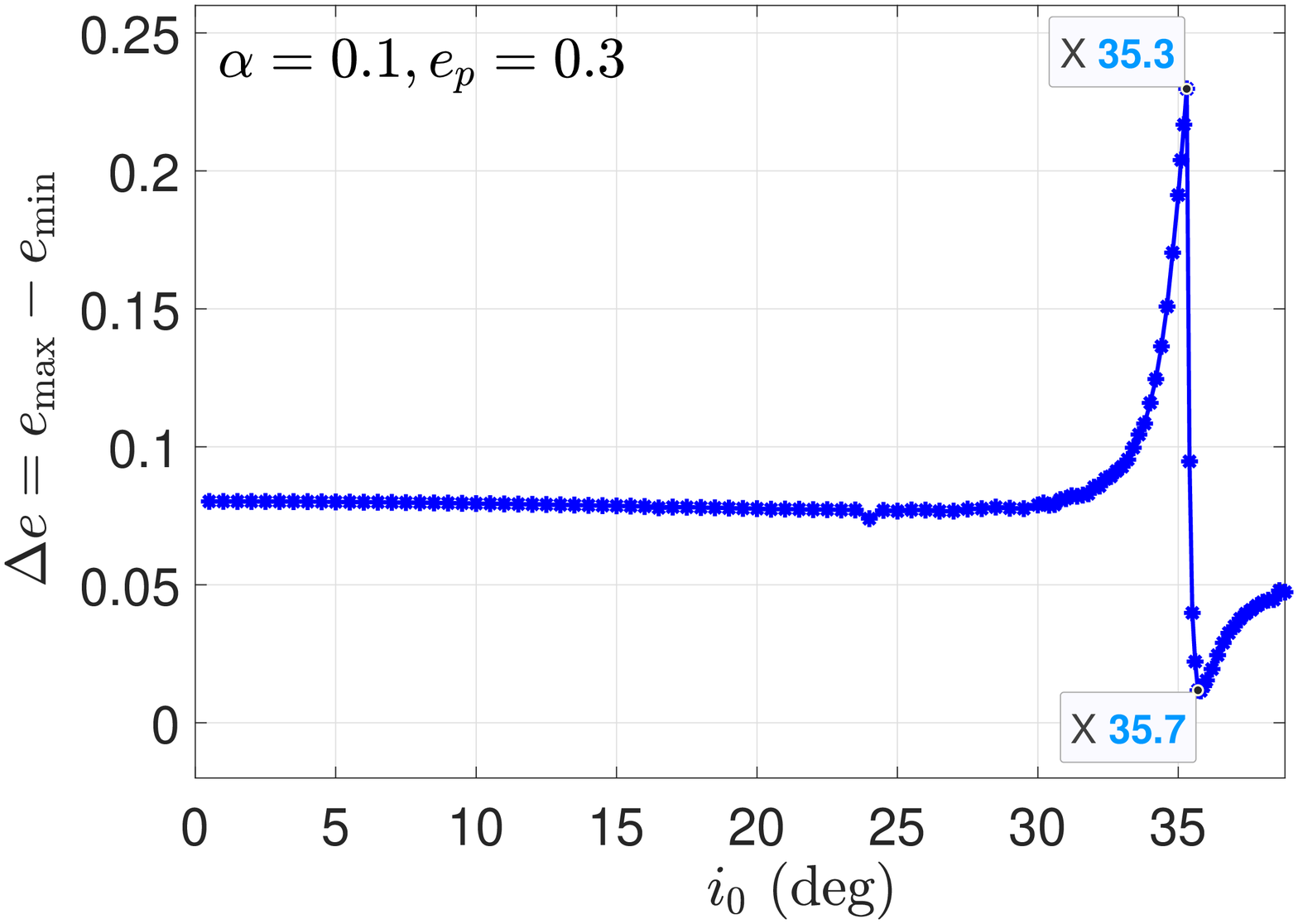}
\includegraphics[width=0.33\textwidth]{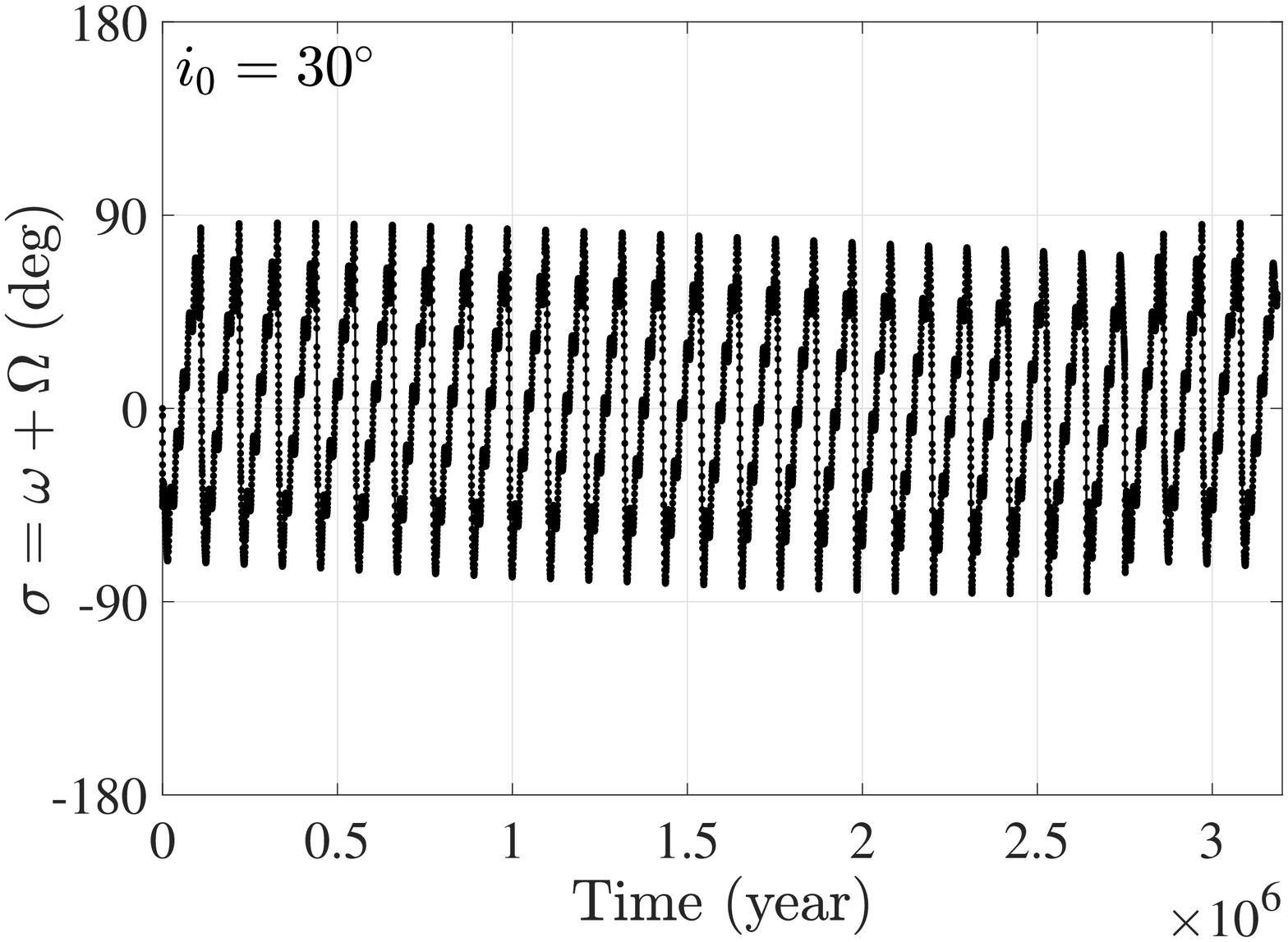}
\includegraphics[width=0.33\textwidth]{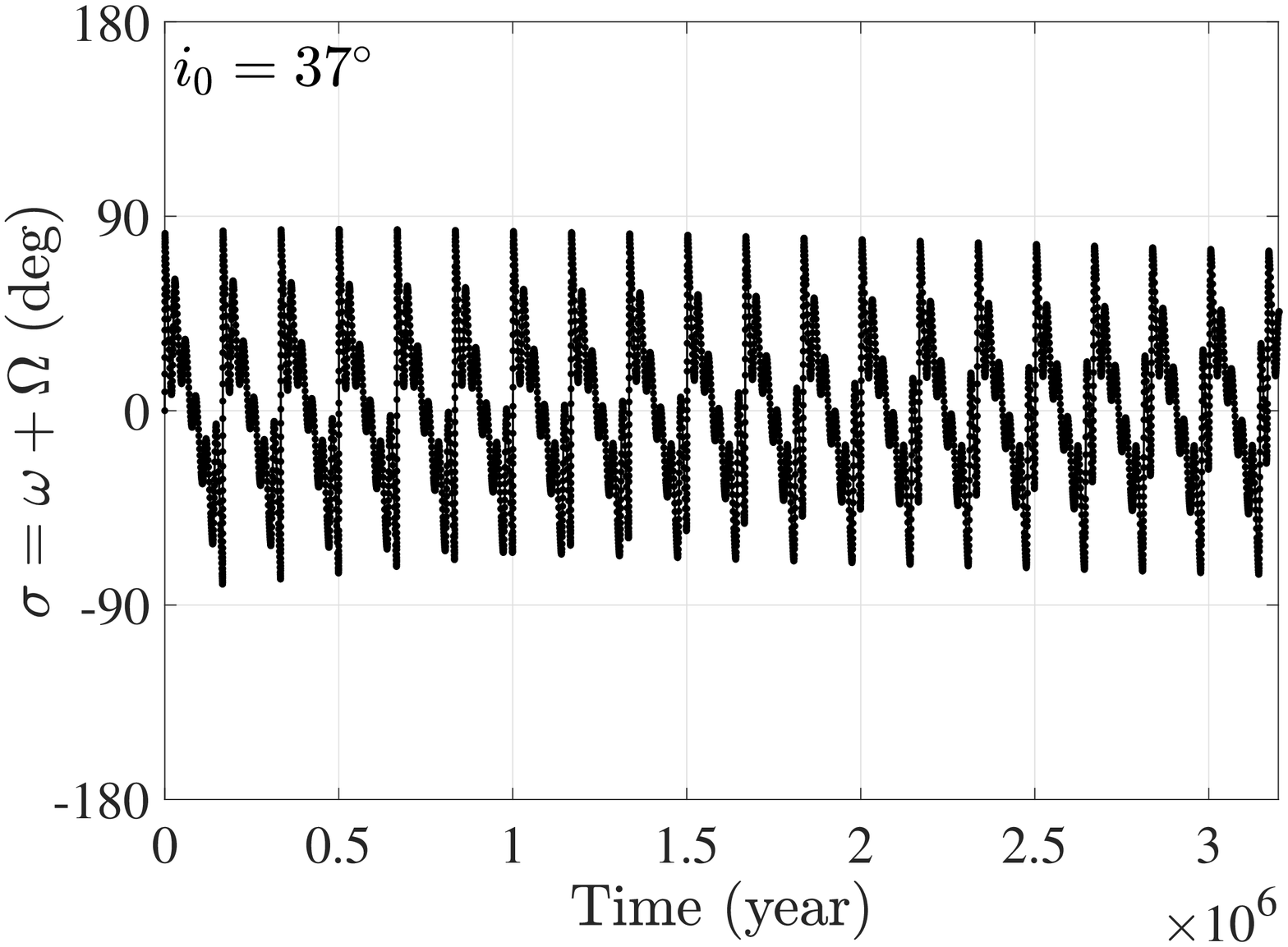}
\caption{The maximum variation of eccentricity as a function of the initial inclination $i_0$, reached during the integration time of $1.6 \times 10^{6}$ $\rm{year}$ (\emph{left panel}) and time histories of the argument $\sigma = \varpi (=\omega + \Omega)$ for inner test particles with $i_0 = 30^{\circ}$ (\emph{middle panel}) and $i_0 = 37^{\circ}$ (\emph{right panel}). The inner test particles are initially placed on quasi-circular orbits ($e_0 = 1 \times 10^{-4}$). The initial conditions of test particles are $a_0 = 0.1$ $\rm{au}$ (i.e., $\alpha = 0.1$), $\Omega_0 = 90^{\circ}$ and $\omega_0 = 270^{\circ}$.}
\label{Fig3}
\end{figure*}

To study the secular dynamics, we perform numerical integrations under the full secular model developed in the previous section and recorded the maximum variation of eccentricity as a function of the initial inclination ranging from $0^{\circ}$ to $39^{\circ}$, as shown in the left panel of Fig. \ref{Fig3}. In the simulations, the inner test particles are placed on quasi-circular orbits ($e_0 = 1 \times 10^{-4}$) and the total integration time is fixed as $1.6 \times 10^{6}$ $\rm{year}$ (it is noted that the integration period taken here is long enough to cover several periods of secular resonances). Please refer to the caption of Fig. \ref{Fig3} for more details about initial conditions. If we change the semimajor axis ratio $\alpha$ or the eccentricity of the giant planet $e_p$, similar plots could be produced (the results are not repeated here). It is noted that the curve of eccentricity variation shown in the left panel of Fig. \ref{Fig3} is in agreement with the results given by \citet{libert2012interesting} (see Fig. 6 in their work where the octupole-order dynamical model was used).

According to the curve of eccentricity variation, we could observe the following dynamical behaviours: (a) a sudden decrease of $\Delta e$ occurs from $i = 35.3^{\circ}$; (b) along the curve of $\Delta e$ there is a deep well located in the inclination interval from $35.3^{\circ}$ to $39^{\circ}$; (c) when the mutual inclination is smaller than $30^{\circ}$, $\Delta e$ seems insensitive to the initial inclination ($\Delta e$ remains $\sim$$0.08$, this value will be explained in Section \ref{Sect5} from the viewpoint of resonant width); (d) when the inclination is greater than $30^{\circ}$ and smaller than $35.3^{\circ}$, $\Delta e$ is a monotonically increasing function of the inclination (the peak of $\Delta e$ could reach $\sim$0.23); (e) when the inclination is greater than $35.8^{\circ}$ and smaller than $38.8^{\circ}$, $\Delta e$ is again an increasing function of inclination and, in particular, it remains relatively small ($\Delta e < 0.05$).

Observing the curve of eccentricity variation shown in the left panel of Fig. \ref{Fig3}, we may ask: what's the dynamical mechanism hidden in the space with inclinations smaller than $39^{\circ}$ for sculpting the shape of $\Delta e$? Our purpose of this work is to provide a possible dynamical explanation for the curve of $\Delta e$.

To provide a preliminary view about this problem, two numerical examples are reported in the middle and right panels of Fig. \ref{Fig3} with initial inclinations at $30^{\circ}$ and $37^{\circ}$ and initial eccentricity at $e_0 = 1 \times 10^{-4}$. It is observed that the argument $\sigma (=\omega + \Omega)$ librates around zero with amplitude smaller than $90^{\circ}$. More numerical simulations still show that the libration of $\sigma = \varpi$ occurs in almost the entire interested region. These preliminary simulations provide us reliable evidence to such a conjecture that the secular dynamics in low-eccentricity region with inclination smaller than $39^{\circ}$ is dominated by the secular resonance associated with $\sigma = \omega + \Omega$. In Section \ref{Sect4}, we will study the secular resonance in detail.

\section{Secular resonances}
\label{Sect4}

In this section, we intend to investigate the secular resonance associated with the critical argument $\sigma = \varpi$ by means of Lie-series perturbation theory and then discuss the resonant dynamics (including the phase-space structures, distribution of resonant centre and width of resonance) in order to provide an explanation for the shape of $\Delta e$ shown in the left panel of Fig. \ref{Fig3}.

\subsection{Perturbation theory}
\label{Sect4-1}

In the current study, the perturbation theory based on Lie-series transformation \citep{hori1966theory, deprit1969canonical}, also known as the Hori--Deprit transformation method, is adopted to eliminate high-order periodic terms from the Hamiltonian in order to formulate secular Hamiltonian.

Before performing the Hori--Deprit transformation, it needs to determine a kernel function that gives the unperturbed frequencies in the secular model. To this end, let us divide the double-averaged Hamiltonian into two parts, denoted by
\begin{equation}\label{Eq6}
{\cal H} = \sum\limits_{{k_1} = 0}^N {\sum\limits_{{k_2 (2)} =  - 2N}^{2N} {{{\cal C}_{{k_1},{k_2}}}\cos \left( {{k_1}p + {k_2}q} \right)} }  = {{\cal H}_0} + \varepsilon {{\cal H}_1},
\end{equation}
where the unperturbed Hamiltonian ${{\cal H}_0} = {\cal C}_{0,0}$ is a function of the action variables $P$ and $Q$, and the perturbed Hamiltonian ${{\cal H}_1}$ is the sum of all those periodic terms, given by
\begin{equation*}
{{\cal H}_1} = \sum\limits_{\left\{ {{k_1},{k_2}} \right\} \backslash \left\{ {0,0} \right\}} {{{\cal C}_{{k_1},{k_2}}}\cos \left( {{k_1}p + {k_2}q} \right)},
\end{equation*}
where $\left\{ {{k_1},{k_2}} \right\} \backslash \left\{ {0,0} \right\}$ means that $k_1$ and $k_2$ are not equal to zero at the same time. Note that the small parameter $\varepsilon$ in equation (\ref{Eq6}) is used to formally separate ${{\cal H}_1}$ from ${{\cal H}_0}$ and it is equal to unity in magnitude. Obviously, the unperturbed Hamiltonian ${{\cal H}_0}$ is totally integrable and thus it can be taken as the kernel function \citep{beauge2006high}, producing the unperturbed frequencies in the following way:
\begin{equation*}
{\upsilon _p} = \frac{{\partial {{\cal H}_0}}}{{\partial P}},\quad {\upsilon _q} = \frac{{\partial {{\cal H}_0}}}{{\partial Q}}.
\end{equation*}
The key concept of Hori--Deprit perturbation theory is searching for a canonical transformation from an old set of variables $(p,q,P,Q)$ to a new set of variables $(p^*,q^*,P^*,Q^*)$, realised by the Lie-series generating function,
\begin{equation*}
S = \sum\limits_k {{\varepsilon ^k}{S_k}\left( {{p^*},{q^*},{P^*},{Q^*}} \right)},
\end{equation*}
in order to satisfy the following identity (the Hamiltonian is conserved before and after the transformation):
\begin{equation}\label{Eq7}
{\cal H} = {{\cal H}_0}\left( {P,Q} \right) + \varepsilon {{\cal H}_1}\left( {p,q,P,Q} \right) = \sum\limits_k {{\varepsilon ^k}{\cal H}_k^*\left( {{p^*},{q^*},{P^*},{Q^*}} \right)},
\end{equation}
where the new Hamiltonian ${\cal H}_k^*$ does not involve those short-period terms.

Applying the Lie-series expansions to the left-hand side of equation (\ref{Eq7}) and equating those terms with equal powers of $\varepsilon$ in equation (\ref{Eq7}), it is possible to obtain the explicit relationship between the new Hamiltonian, generating function and the old Hamiltonian (up to the second order in $\varepsilon$ as an example) as follows \citep{hori1966theory}:
\begin{equation}\label{Eq8}
\begin{aligned}
{\rm Order\;0}:\quad {\cal H}_0^* &= {{\cal H}_0},\\
{\rm Order\;1}:\quad {\cal H}_1^* &= {{\cal H}_1} + \left\{ {{{\cal H}_0},{S_1}} \right\},\\
{\rm Order\;2}:\quad {\cal H}_2^* &= {{\cal H}_2} + \left\{ {{{\cal H}_0},{S_2}} \right\} + \frac{1}{2}\left\{ {{{\cal H}_1} + {\cal H}_1^*,{S_1}} \right\},
\end{aligned}
\end{equation}
where both the left- and right-hand terms are evaluated at the new canonical variables (i.e. the variables with superscript of star), and $\left\{*, *\right\}$ denotes the Poisson bracket.

\subsection{Resonant Hamiltonian}
\label{Sect4-2}

We will use equation (\ref{Eq8}) to formulate resonant Hamiltonian up to the second order in $\varepsilon$.

If a test particle is located inside the resonant region, the critical argument $\sigma = p (=\varpi)$ becomes the `long-period' variable and the angle $q$ is the `short-period' variable. Thus, in the double-averaged Hamiltonian, the components involving $q$ are short-period terms and the remaining components are long-period or secular terms. In order to study the resonant dynamics, it is usual to eliminate those short-period terms from the Hamiltonian .

As discussed in Section \ref{Sect4}, the elimination of high-order short-period effects can be realised through the Hori--Deprit method by using equation (\ref{Eq8}). At the zeroth order in $\varepsilon$, it is obvious that the new Hamiltonian is equal to the old one,
\begin{equation*}
{\cal H}_0^* = {{\cal H}_0}.
\end{equation*}
At the first order in $\varepsilon$, we have
\begin{equation}\label{Eq10}
{\cal H}_1^* = {{\cal H}_1} + \left\{ {{{\cal H}_0},{S_1}} \right\},
\end{equation}
where ${\cal H}_1^*$ is the long-term part of the right-hand side, given by
\begin{equation}\label{Eq10-1}
{\cal H}_1^* = \sum\limits_{k = 1}^N {{{\cal C}_{k,0}}\cos \left( {k\sigma} \right)}
\end{equation}
and the periodic part of equation (\ref{Eq10}) related to the generating function $S_1$ holds
\begin{equation}\label{Eq11}
- \left\{ {{{\cal H}_0},{S_1}} \right\} = {{\cal H}_1} - {\cal H}_1^* = \sum\limits_{{k_2} \ne 0} {{{\cal C}_{{k_1},{k_2}}}\cos \left( {{k_1}p^* + {k_2}q^*} \right)},
\end{equation}
where $k_1$ changes from $0$ to $N$ and $k_2$ from $-2N$ to $2N$ with step 2. It should be noted that, in equation (\ref{Eq10-1}), $\sigma$ is evaluated at the starred variables, i.e. $\sigma = p^*$. As ${\cal H}_0$ is the kernel function of the problem at hand, we have
\begin{equation*}
- \left\{ {{{\cal H}_0},{S_1}} \right\} = {\upsilon _p}\frac{{\partial {S_1}}}{{\partial p^*}} + {\upsilon _q}\frac{{\partial {S_1}}}{{\partial q^*}},
\end{equation*}
so that the first-order generating function can be obtained as
\begin{equation}\label{Eq12}
{S_1} = \sum\limits_{{k_2} \ne 0} {\frac{{{{\cal C}_{{k_1},{k_2}}}}}{{{k_1}{\upsilon _p} + {k_2}{\upsilon _q}}}\sin \left( {{k_1}p^* + {k_2}q^*} \right)}.
\end{equation}
Extending to the second order in $\varepsilon$, we have the following equality,
\begin{equation}\label{Eq13}
{\cal H}_2^* = \left\{ {{{\cal H}_0},{S_2}} \right\} + \frac{1}{2}\left\{ {{{\cal H}_1} + {\cal H}_1^*,{S_1}} \right\}.
\end{equation}
Similar to the previous step, it is possible to obtain the second-order resonant Hamiltonian ${\cal H}_2^*$ which is composed of the secular terms of the right-hand side (i.e., removing those short-period terms in the right-hand side to obtain ${\cal H}_2^*$). For the current problem at hand, we only perform this process up to the second order in $\varepsilon$. Theoretically speaking, it could be formulated up to an arbitrary order.

In summary, the resonant Hamiltonian, up to the second order in $\varepsilon$, can be expressed as
\begin{equation}\label{Eq18}
{{\cal H}^*} = {\cal H}_0^* + \varepsilon {\cal H}_1^* + {\varepsilon}^2{\cal H}_2^*.
\end{equation}
It is obvious that the angle variable $q$ disappears from the resonant Hamiltonian, meaning that its conjugate variable $Q$ is a conserved quantity in the resonant model (i.e., $Q$ is the motion integral of the resonant model). As a result, the system specified by ${{\cal H}^*}$ reduces to a dynamical model with a single degree of freedom, which is integrable. Based on the resonant Hamiltonian given by equation (\ref{Eq18}), the phase-space structures in the space $(\sigma, P)$ can be produced by plotting level curves of ${\cal H}^*$ with a given motion integral $Q$. According to the phase-space structures, it is possible for us to identify the location of resonant centre and resonant width.

For convenience, we adopt the minimum value of inclination, denoted by $i_{\min}$, to stand for $Q$ by means of $Q = \sqrt{\mu a} (\cos{i_{\min}} - 1)$ (i.e., $i_{\min}$ is the magnitude of inclination when the eccentricity is assumed at zero).

\subsection{Resonant centre and width}
\label{Sect4-3}

As formulated in the previous subsection, the resonant Hamiltonian formulated up to the second order in $\varepsilon$ can be organised in the Fourier-series form:
\begin{equation}\label{Eq19}
{{\cal H}^{{*}}}\left({Q}; {\sigma ,P} \right) = {\cal H}_0^* + \varepsilon {\cal H}_1^* + {\varepsilon^2} {\cal H}_2^* = \sum\limits_{k = 0}^{2N} {{{\cal D}_k}\left( {P ,Q} \right)\cos \left( {k\sigma } \right)},
\end{equation}
where ${\cal D}_k$ are coefficients related to the action variables $P$ and $Q$ (here $Q$ is the motion integral in the resonant model). The equations of motion for the resonant model can be written as
\begin{equation}\label{Eq20}
\begin{aligned}
\dot \sigma  &= \frac{{\partial {{\cal H}^{\rm{*}}}}}{{\partial P }} = \sum\limits_{k = 0}^{2N} {\frac{{\partial {{\cal D}_k}\left( {P ,Q} \right)}}{{\partial P }}\cos \left( {k\sigma } \right)},\\
\dot P  &=  - \frac{{\partial {{\cal H}^{\rm{*}}}}}{{\partial \sigma }} = \sum\limits_{k = 0}^{2N} {k{{\cal D}_k}\left( {P , Q} \right)\sin \left( {k\sigma } \right)},
\end{aligned}
\end{equation}
where the stationary solution (also called equilibrium point) is determined by
\begin{equation}\label{Eq20-1}
\dot{\sigma}=\frac{{\partial {{\cal H}^{\rm{*}}}}}{{\partial P }} =0,\quad \dot{P} = -\frac{{\partial {{\cal H}^{\rm{*}}}}}{{\partial \sigma }} = 0,
\end{equation}
meaning that the resonant Hamiltonian reaches its extremum at an equilibrium point. Putting equation (\ref{Eq20}) into the stationary condition shown by equation (\ref{Eq20-1}), we can get that the equilibrium points occur at $\sigma = 0$ or $\sigma = \pi$.

Linearizing the equations of motion given by equation (\ref{Eq20}) in the vicinity of an equilibrium point yields a linear system and the eigenvalues of its coefficient matrix determines the stability of the equilibrium point. The stability condition requires
\begin{equation*}
{\left( {\frac{{{\partial ^2}{{\cal H}^{\rm{*}}}}}{{\partial \sigma \partial P }}} \right)^2} < \frac{{{\partial ^2}{{\cal H}^{\rm{*}}}}}{{\partial {\sigma ^2}}}\frac{{{\partial ^2}{{\cal H}^{{*}}}}}{{\partial {P ^2}}},
\end{equation*}
where both the left- and right-hand sides are evaluated at the considered equilibrium point. Usually, the stable equilibrium corresponds to an elliptic point which is the resonant centre in the resonant model and the unstable one is called a saddle point. The level curves of Hamiltonian passing through the saddle point correspond to the dynamical separatrices, dividing the entire phase space into libration and circulation regions.

In the phase portrait specified by a given $i_{\min}$ (equivalently, the motion integral $Q$ is given), let's denote the stable solution as $(\sigma_s, P_s)$ and the unstable solution as $(\sigma_u, P_u)$. Usually, the level curves of resonant Hamiltonian passing through the unstable point corresponds to dynamical separatrices. The Hamiltonian of the dynamical separatrix is given by
\begin{equation*}
{{\cal H}^{\rm{*}}}\left( {{Q};{\sigma _u},{P _u}} \right) = {{\cal H}^{\rm{*}}}\left( {{Q};{\sigma _s},{P _1}} \right) = {{\cal H}^{\rm{*}}}\left( {{Q};{\sigma _s},{P _2}} \right)
\end{equation*}
and the distance between the separatrices evaluated at $\sigma = \sigma_s$ defines the resonant width, calculated by $\Delta P = P_2 - P_1$. For convenience, the resonant width is usually measured in terms of the variation of eccentricity, given by
\begin{equation}\label{Eq20-2}
\Delta e =  - \frac{{\Delta P }}{\sqrt{\mu a}}\frac{1}{{\sqrt {{{\left( {\frac{\sqrt{\mu a}}{{{P_s + \sqrt{\mu a}}}}} \right)}^2} - 1}}},
\end{equation}
which is used to determine the resonant width in the following simulations. Please see Figs. \ref{Fig5}--\ref{Fig6} for the definition of $\Delta e$ in phase portraits.

\section{Results}
\label{Sect5}

In this section, the analytical developments developed in Section \ref{Sect4} are applied to secular resonances occurring in the interested parameter space (i.e., the low-eccentricity region with inclinations smaller than $39^{\circ}$). Initially, the phase portraits can be produced by plotting level curves of the resonant Hamiltonian. By analyzing the structures arising in phase portraits, it is possible to identify the libration zones and measure the size of libration islands (i.e., width of resonance). At last, a correspondence is made between the resonant width and the maximum variation of eccentricity $\Delta e$, which provides an explanation for the shape of $\Delta e$ numerically determined in section \ref{Sect3}.

\subsection{Phase portraits}
\label{Sect5-1}

\begin{figure*}
\centering
\includegraphics[width=0.45\textwidth]{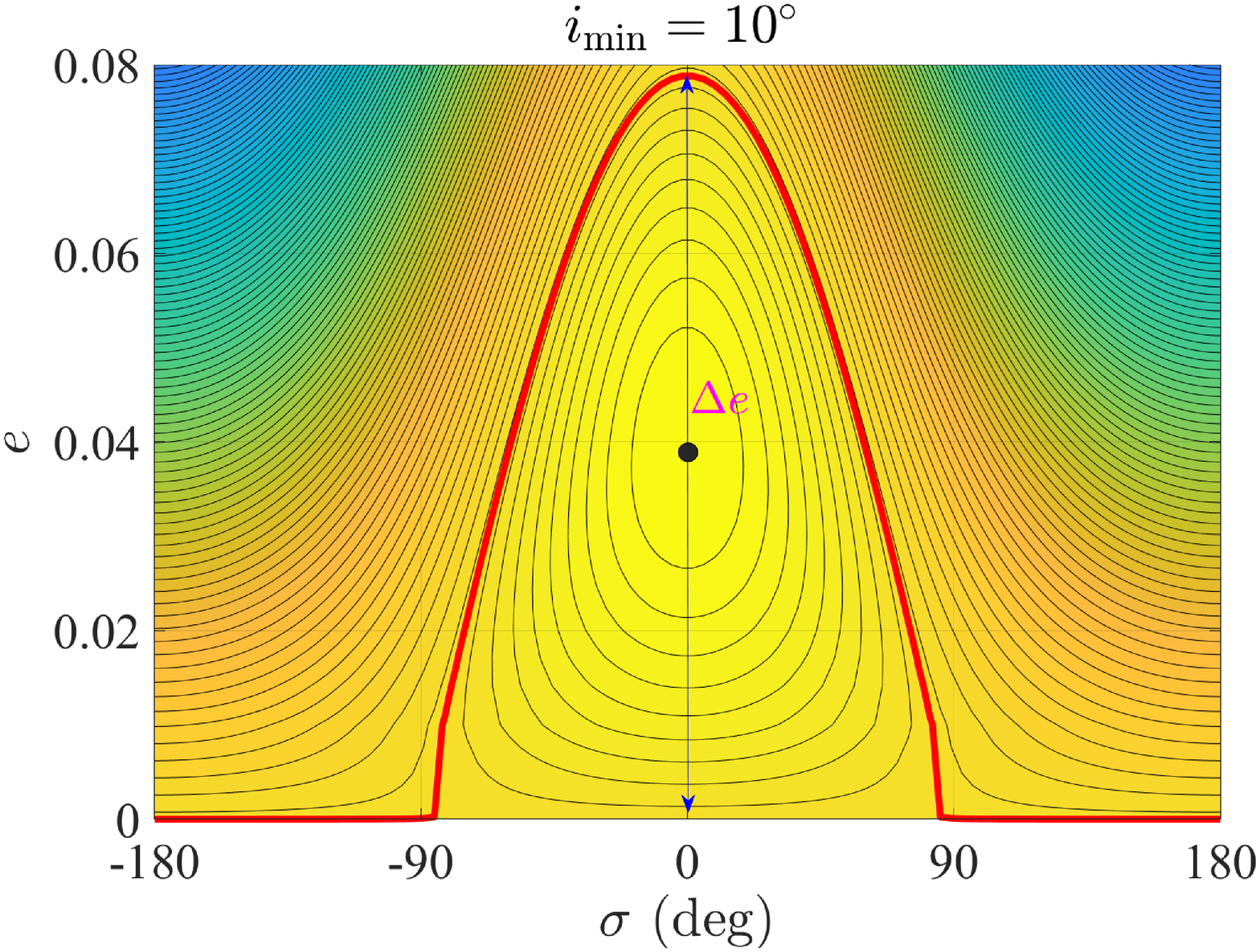}
\includegraphics[width=0.45\textwidth]{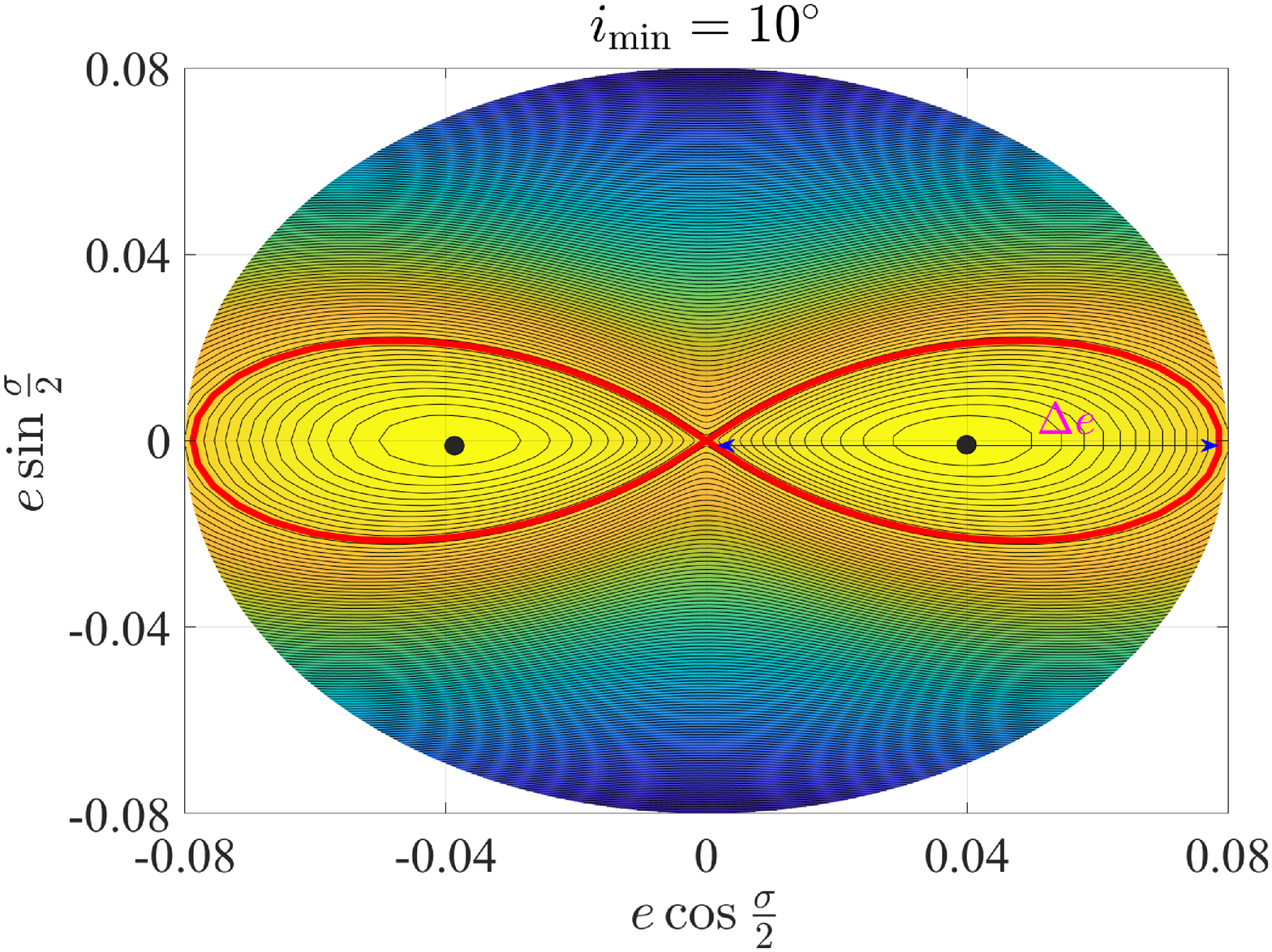}
\caption{Level curves of the resonant Hamiltonian ${\cal H}^*$ specified by the minimum value of inclination at $i_{\min} = 10^{\circ}$. In simulations, the semimajor axis ratio is taken as $\alpha = 0.1$. The resonant centres are marked by black dots and the level curves separating the libration regions from circulation regions are shown in red lines. The resonant width is measured by the variation of eccentricity, denoted by $\Delta e$.}
\label{Fig5}
\end{figure*}

\begin{figure*}
\centering
\includegraphics[width=0.45\textwidth]{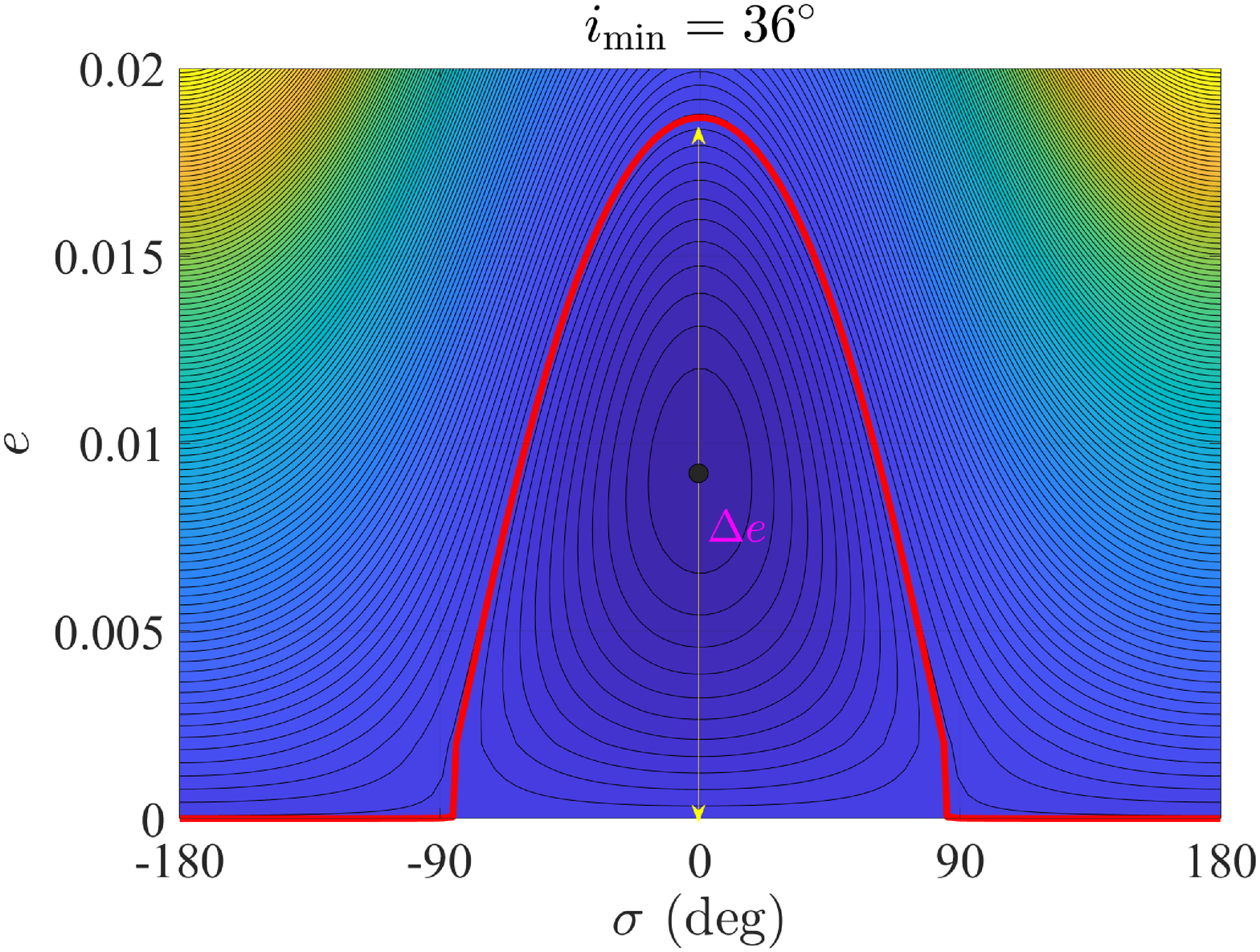}
\includegraphics[width=0.45\textwidth]{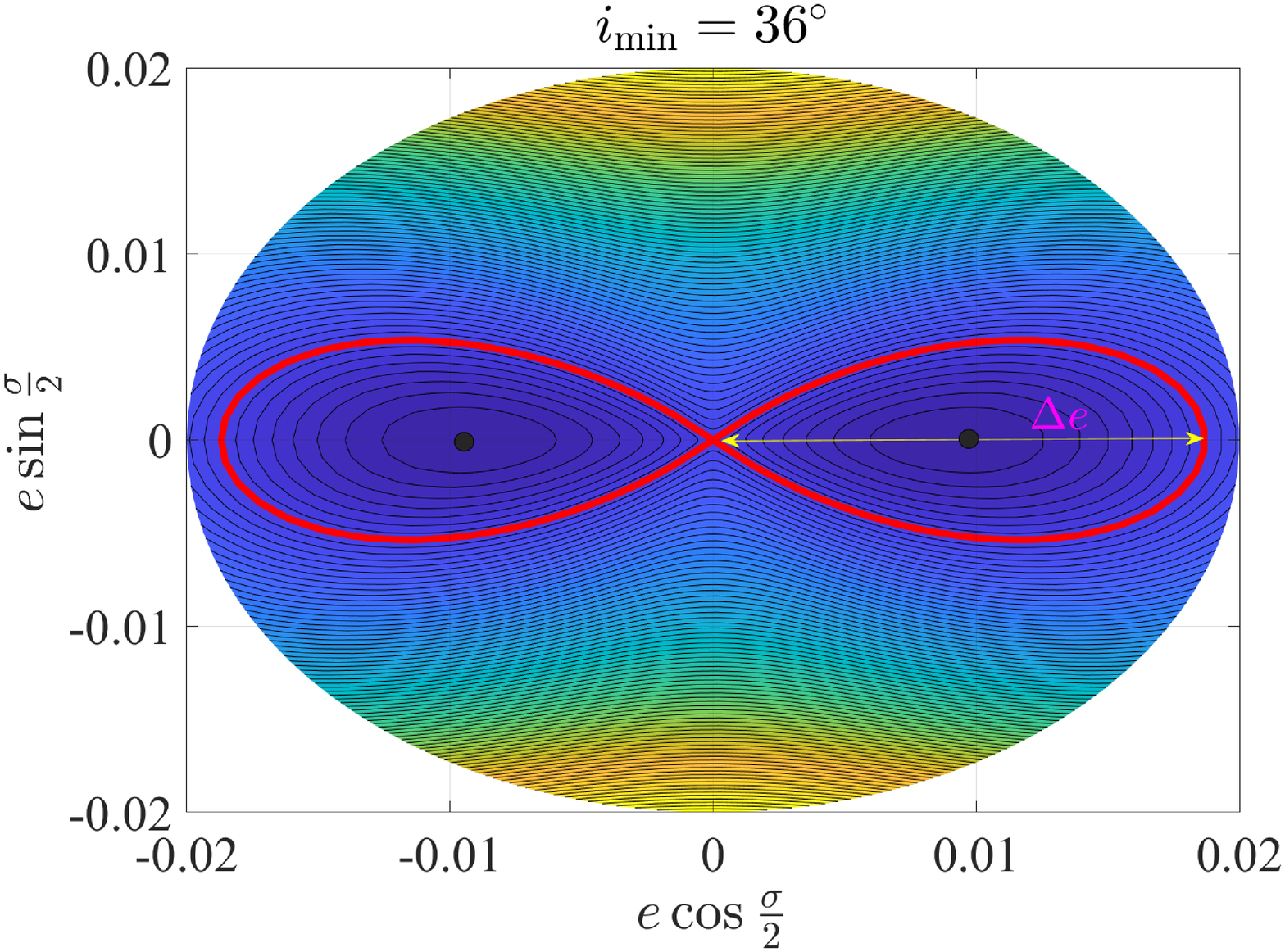}
\caption{Similar to Fig. \ref{Fig5} but for the resonant Hamiltonian ${\cal H}^*$ specified by $i_{\min} = 36^{\circ}$.}
\label{Fig6}
\end{figure*}

As examples, we take the minimum values of inclination at $i_{\min} = 10^{\circ}$ and $i_{\min} = 36^{\circ}$ to show the phase portraits. In this work, we adopt the polar coordinate $(e\cos{\sigma/2}, e\sin{\sigma/2})$ instead of $(e\cos{\sigma}, e\sin{\sigma})$ to show phase portraits due to the consideration that the zero-eccentricity point (i.e., the coordinate origin) is a visible saddle point in the $(e\cos{\sigma/2}, e\sin{\sigma/2})$ space.

The level curves of resonant Hamiltonian (i.e., phase portraits) specified by $i_{\min} = 10^{\circ}$ are reported in Fig. \ref{Fig5}, in which the resonant centres are shown by black dots and the level curves which separate the libration regions from the circulation regions are marked in red lines. From Fig. \ref{Fig5}, it is observed that (a) the zero-eccentricity point (i.e., the coordinate origin in the polar coordinate plane) is a visible saddle point in the resonant model, (b) the resonant centre is located at $\sigma = 0^{\circ}$, (c) there are two types of motion modes, including rotation and libration, which are separated by the level curve stemming from zero-eccentricity point (it corresponds to the separatrix in the resonant model), (d) along the line of separatrix the eccentricity could be excited from $\sim$0 up to $\sim$$0.08$ (remind that this value is approximately equal to the value of $\Delta e$ numerically determined in Section \ref{Sect3}), and (e) the distance between the separatrix in terms of the variation of eccentricity defines the resonant width, demoted by $\Delta e$, as shown in Fig. \ref{Fig5}.

The phase portraits of the secular resonance specified by $i_{\min} = 36^{\circ}$ are shown in Fig. \ref{Fig6}, which shows that (a) the resonant centre is located at $\sigma = 0^{\circ}$ and (b) the level curve stemming from zero eccentricity divides the entire phase space into regions of rotation and libration types. The distance between the separatrices evaluated at the resonant centre also defines the resonant width, denoted by $\Delta e$.

\begin{figure*}
\centering
\includegraphics[width=0.45\textwidth]{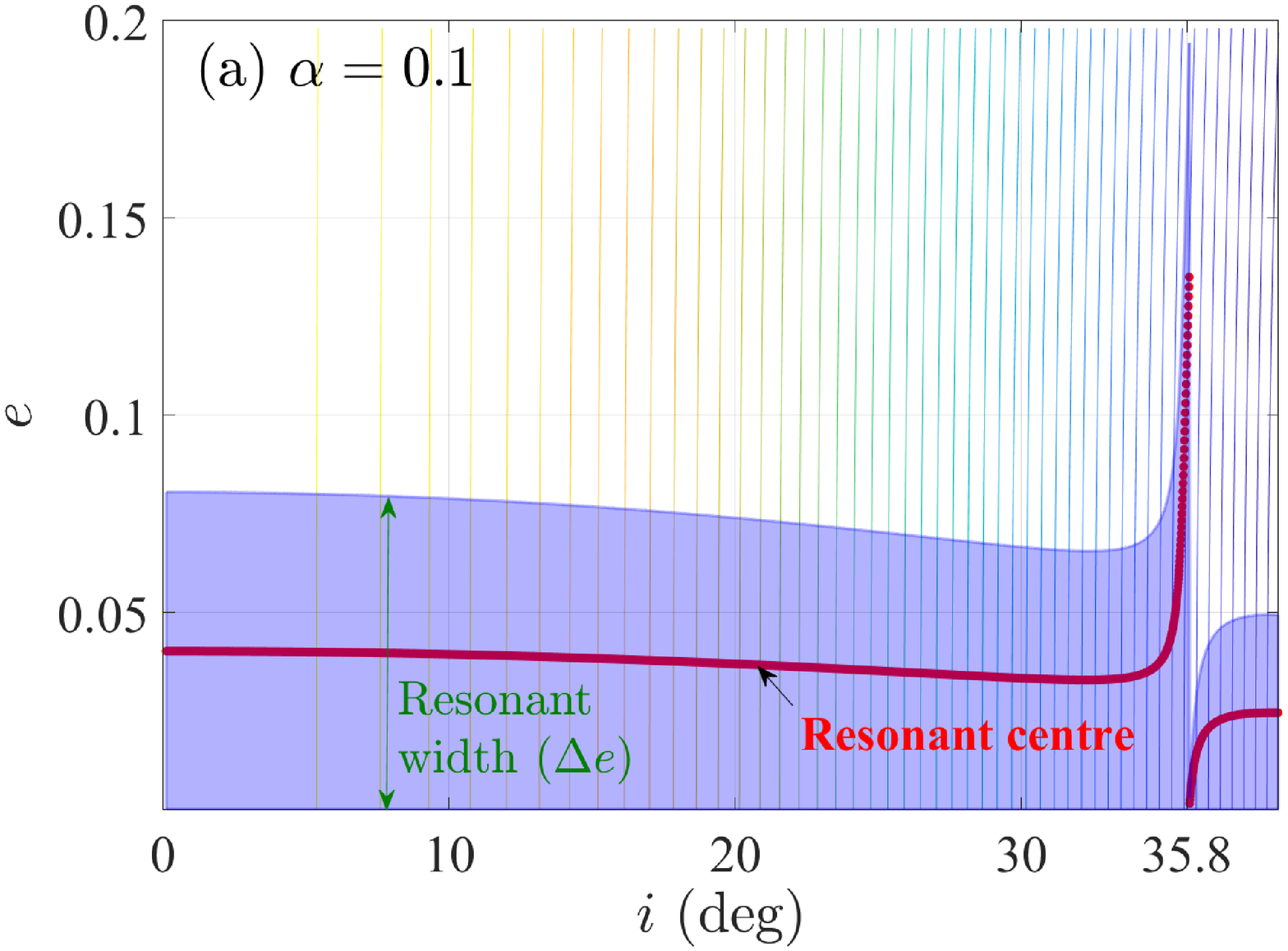}
\includegraphics[width=0.45\textwidth]{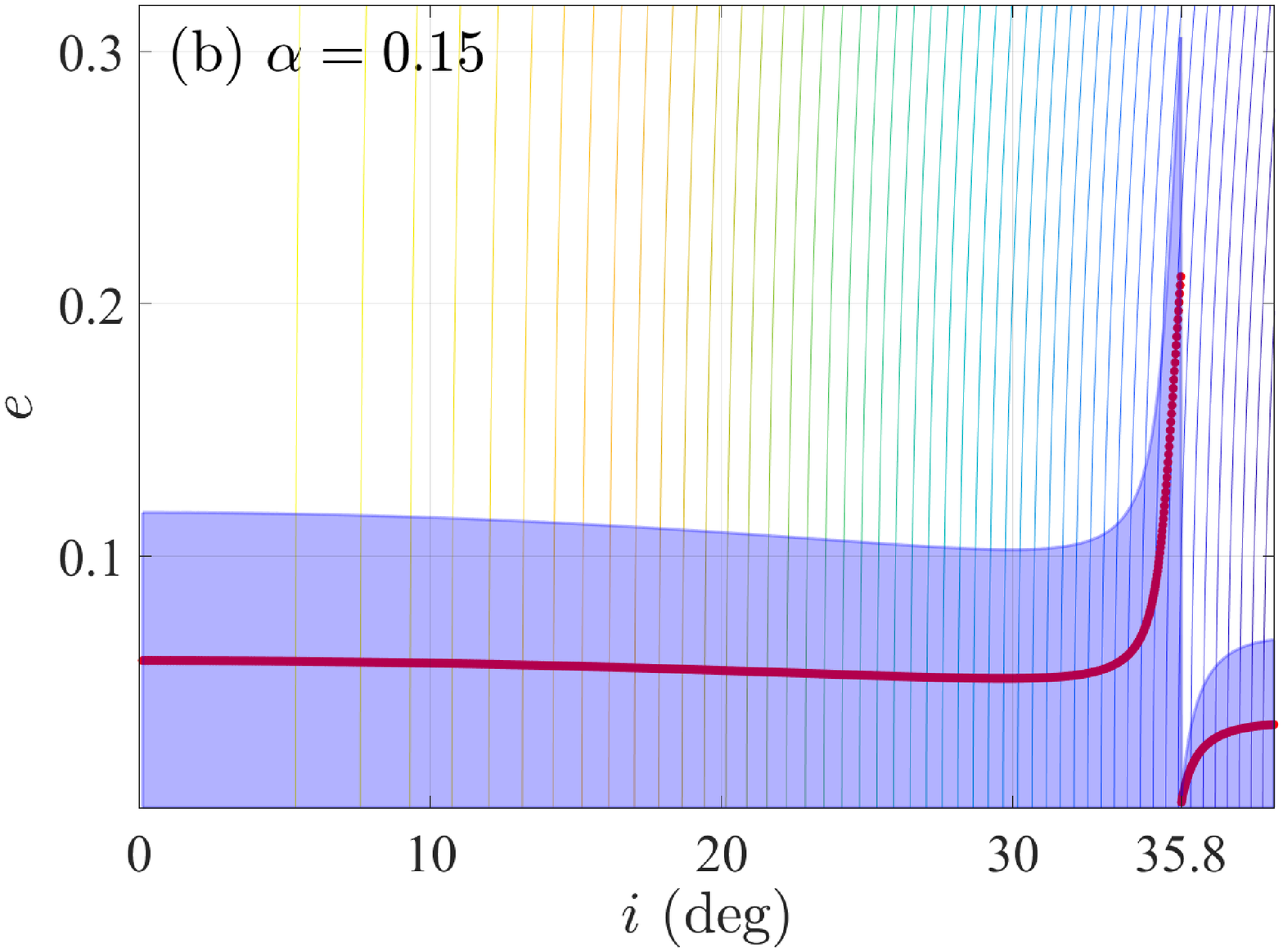}
\includegraphics[width=0.45\textwidth]{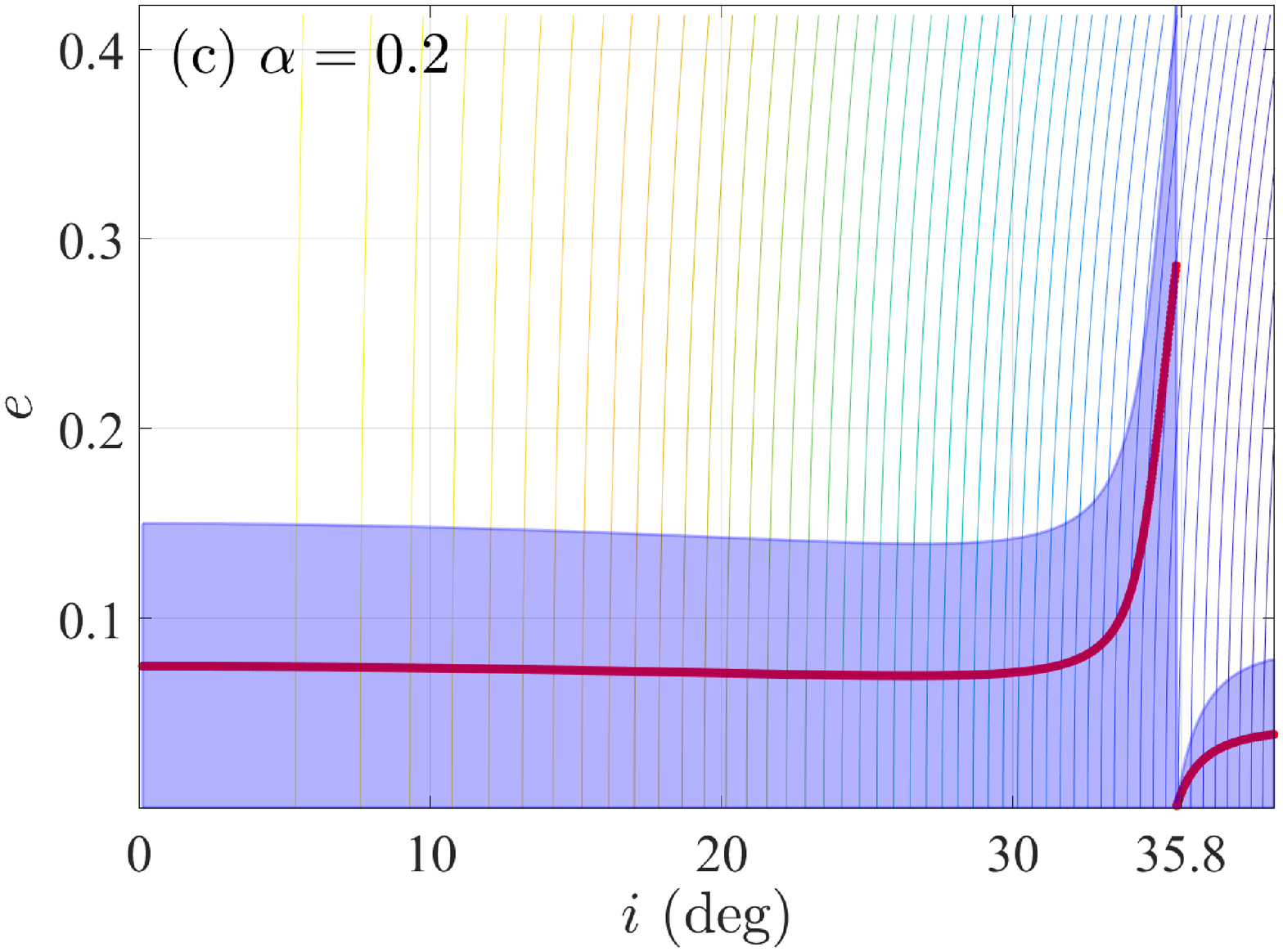}
\includegraphics[width=0.45\textwidth]{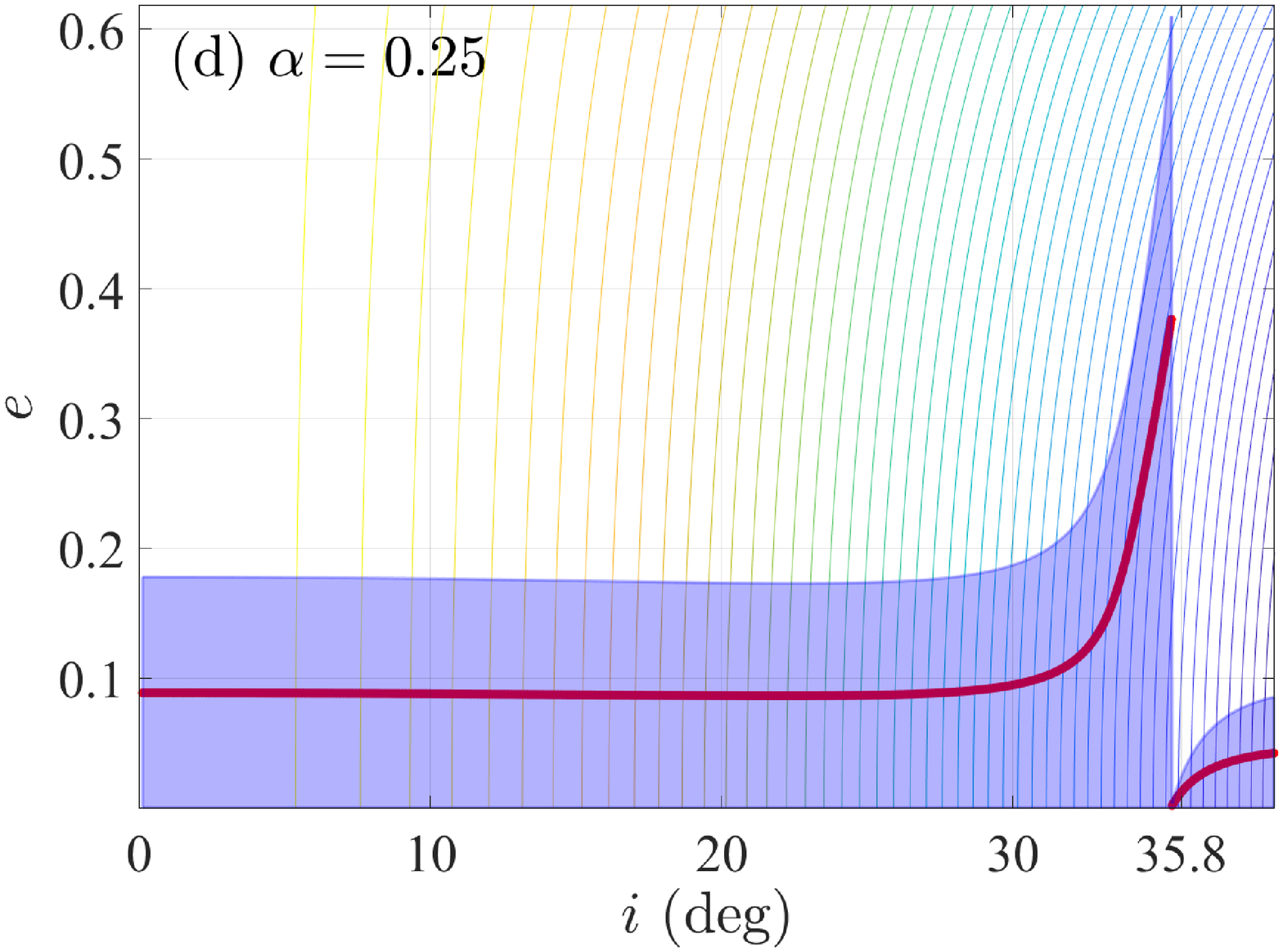}
\caption{Analytical results of resonant centre and resonant width in terms of $\Delta e$ for different semimajor axis ratios at $\alpha = 0.1$, $\alpha=0.15$, $\alpha=0.2$ and $\alpha=0.25$. The resonant width is denoted by $\Delta e$, as shown in the first panel. For convenience, the level curves of the motion integral $Q = \sqrt{\mu a (1-e^2)} (\cos{i} - 1)$ are also plotted (the width of resonance should be measured along the level curve of $Q$).}
\label{Fig7}
\end{figure*}

\subsection{Resonant width and explanation for the curve of $\Delta e$}
\label{Sect5-2}

According to the definition of resonant width, we can see that, the initially quasi-circular orbits are close to the saddle point (zero-eccentricity point) in the phase space, meaning that the real trajectories of test particles should be close to the curve of separatrix, so that the maximum variation of eccentricity numerically determined in Section \ref{Sect3} can be estimated by the associated resonant width.

Our main results are reported in Fig. \ref{Fig7}, where the location of resonant centre and the resonant width in terms of the eccentricity variation $\Delta e$ are reported for four different semimajor axis ratios at $\alpha = 0.1,0.15,0.2,0.25$. It is observed from Fig. \ref{Fig7} that (a) the resonant width evaluated at a given inclination increases with the semimajor axis ratio, (b) there is an interrupt for the curve of resonant width, occurring at inclination of $\sim$$35.8^{\circ}$, and (c) the resonant width, denoted by $\Delta e$, is first a decreasing function and then an increasing function in the inclination interval smaller than $\sim$$35.8^{\circ}$, and it is again an increasing function of inclination when the inclination changes from $\sim$$35.8^{\circ}$ to $39^{\circ}$ .

Comparing the first panel of Fig. \ref{Fig7} with the curve of $\Delta e$ shown in the left panel of Fig. \ref{Fig3} (both plots have the same semimajor axis ratio at $\alpha=0.1$), we can find that the curve of $\Delta e$ is in quite good agreement with the curve of resonant width in terms of the following points: (a) in both plots the point of discontinuity happens at inclination $\sim$$35.8^{\circ}$, which divides the entire interval of inclination into two branches, and (b) the maximum variation of eccentricity determined in Section \ref{Sect3} and the resonant width obtained by the analytical method have similar behaviour as the inclination changing from zero to $39^{\circ}$. Thus, we could conclude that the shape of $\Delta e$ numerically produced in Section \ref{Sect3} is mainly sculpted by the secular resonance associated with $\sigma = \varpi$. In other words, the secular dynamics of inner test particles in the low-eccentricity space with inclinations smaller than $39^{\circ}$ is dominantly governed by the secular resonance with $\sigma = \varpi$.

\subsection{Discussions on the dynamics around the critical inclination}
\label{Sect5-3}

\begin{figure*}
\centering
\includegraphics[width=0.45\textwidth]{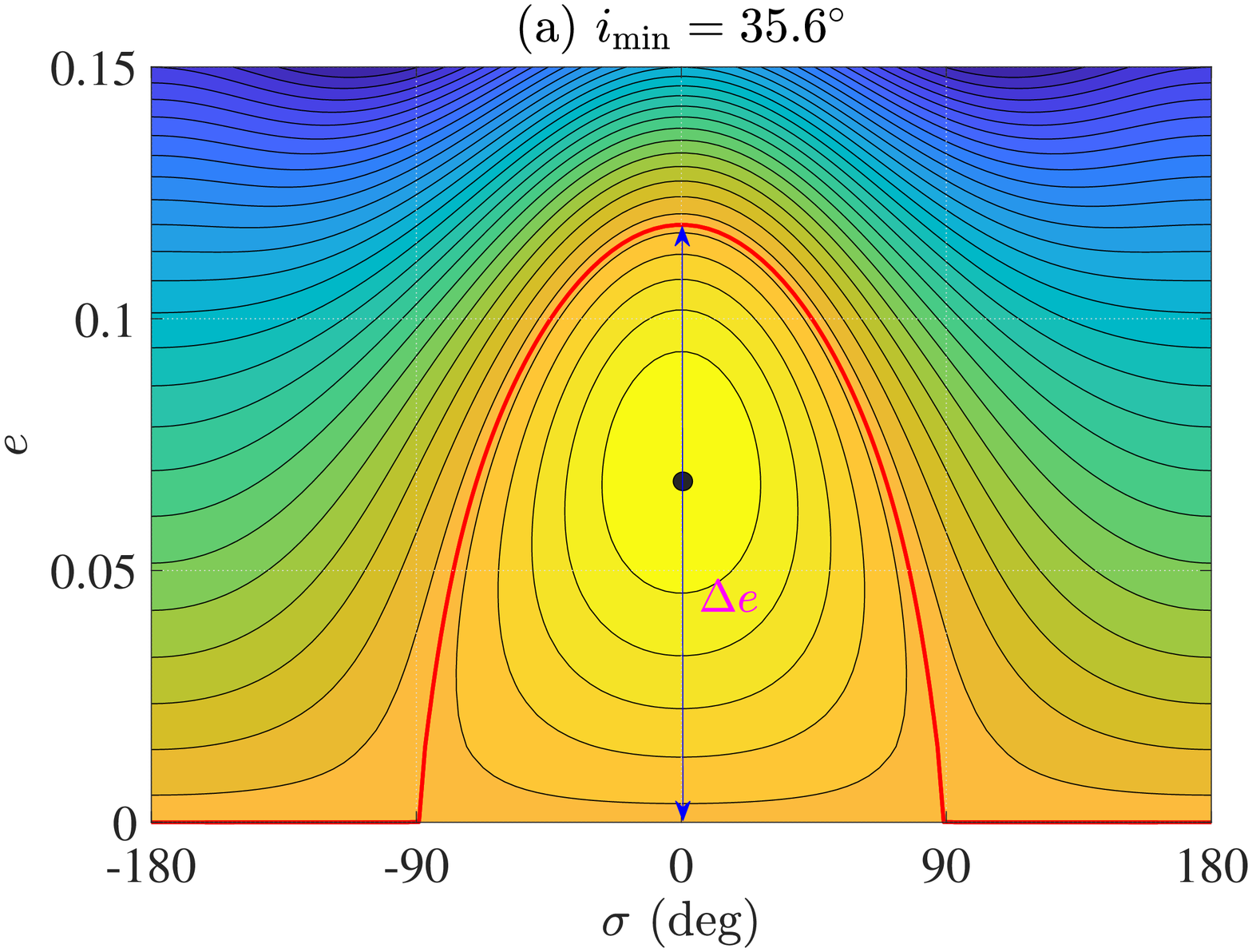}
\includegraphics[width=0.45\textwidth]{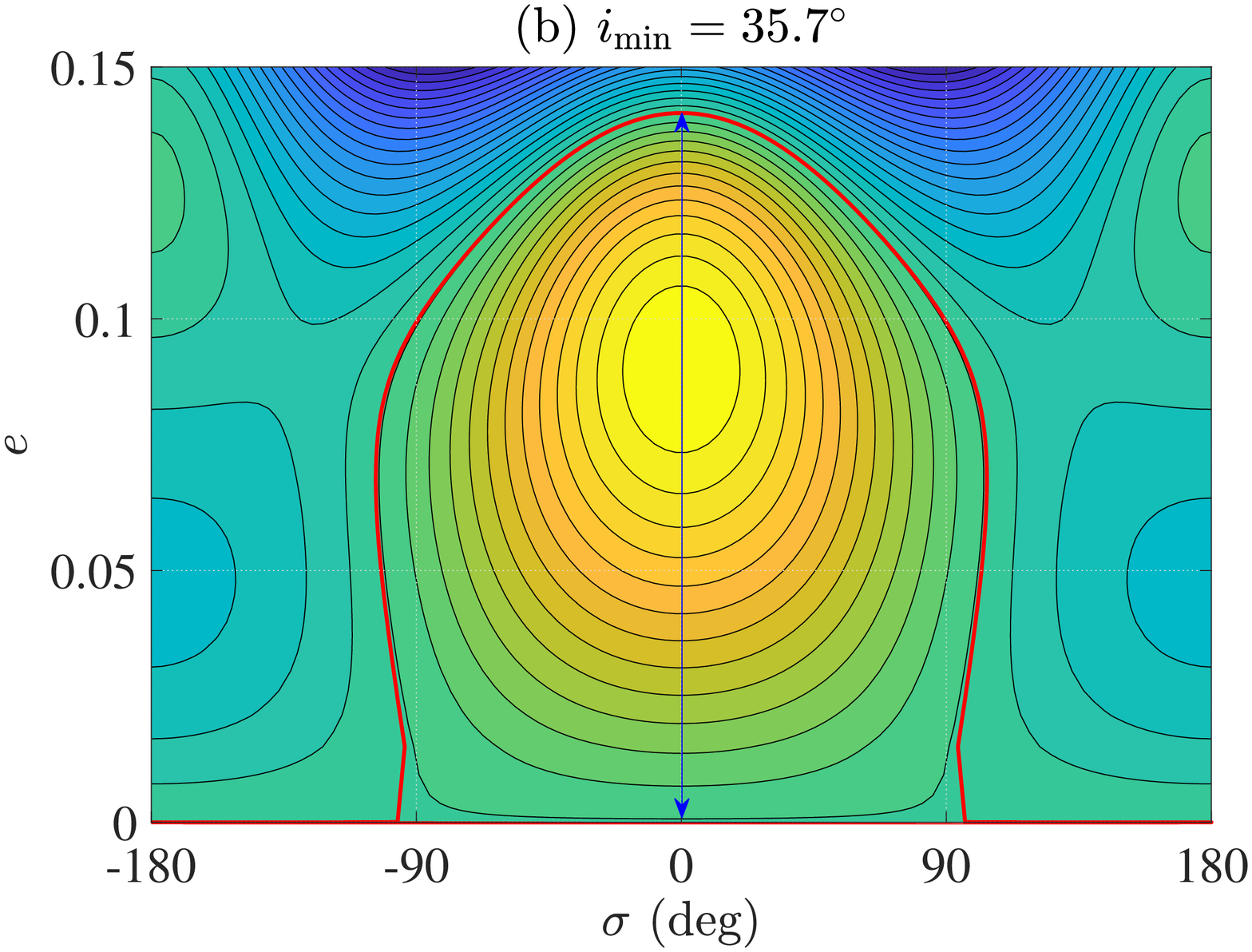}\\
\includegraphics[width=0.45\textwidth]{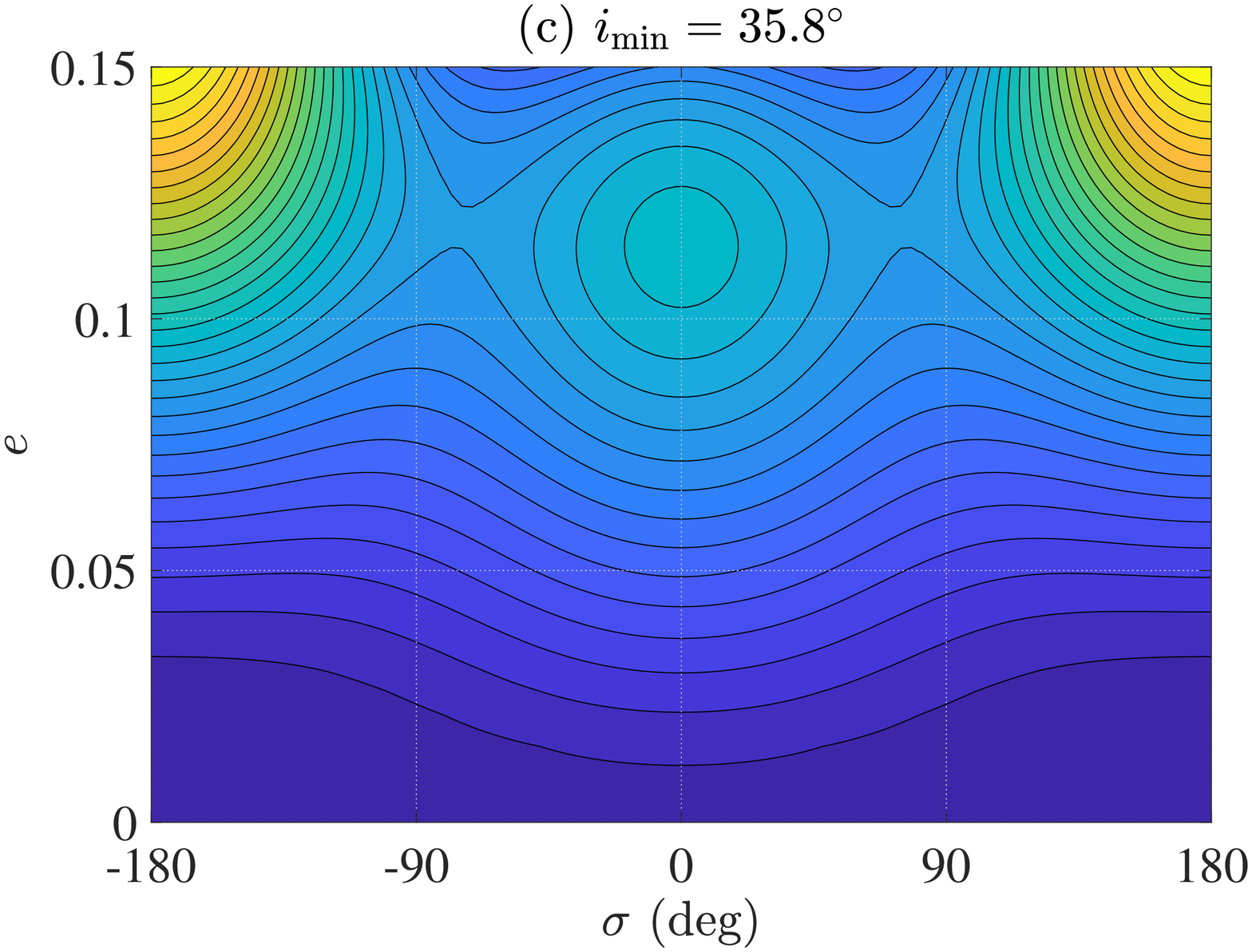}
\includegraphics[width=0.45\textwidth]{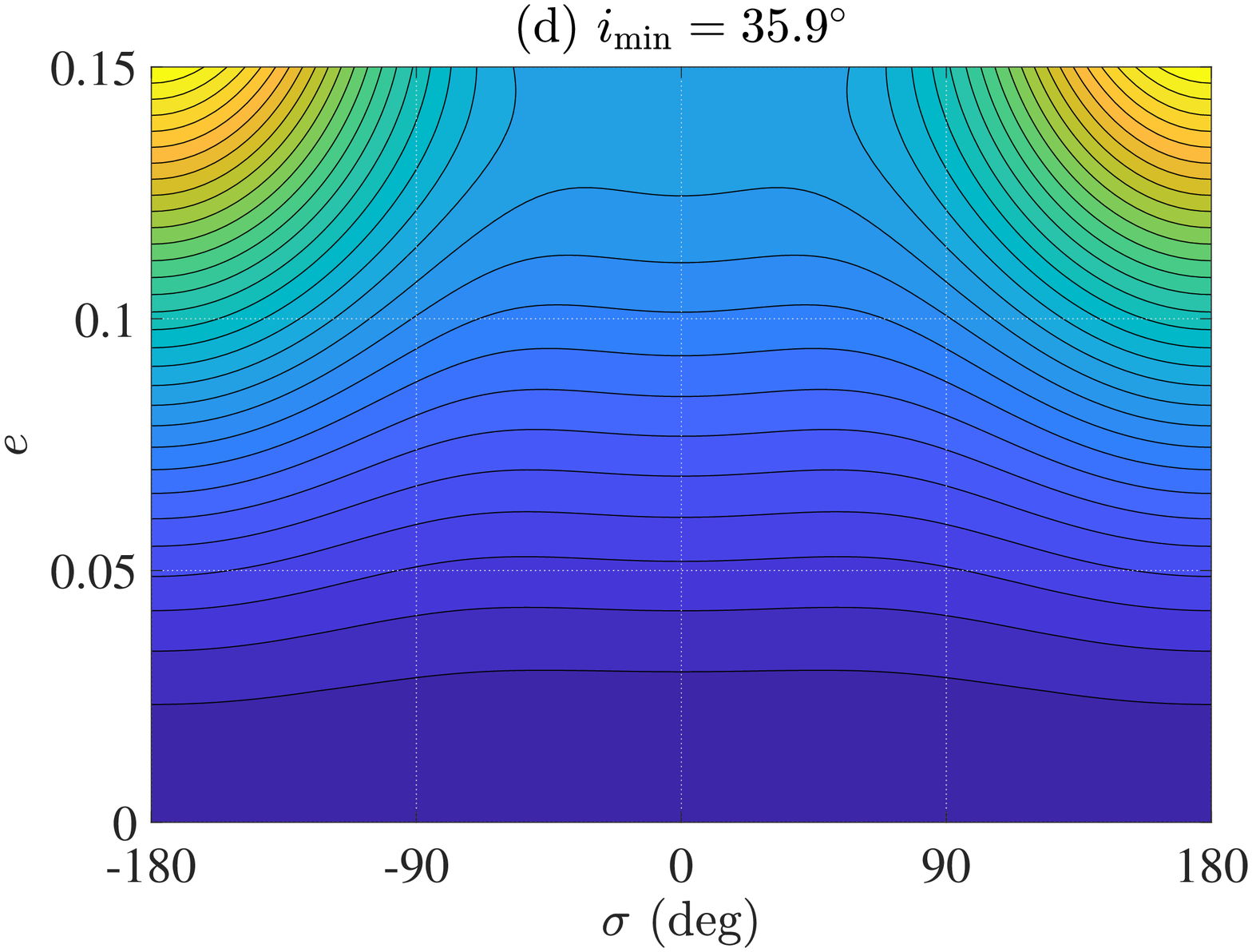}\\
\includegraphics[width=0.45\textwidth]{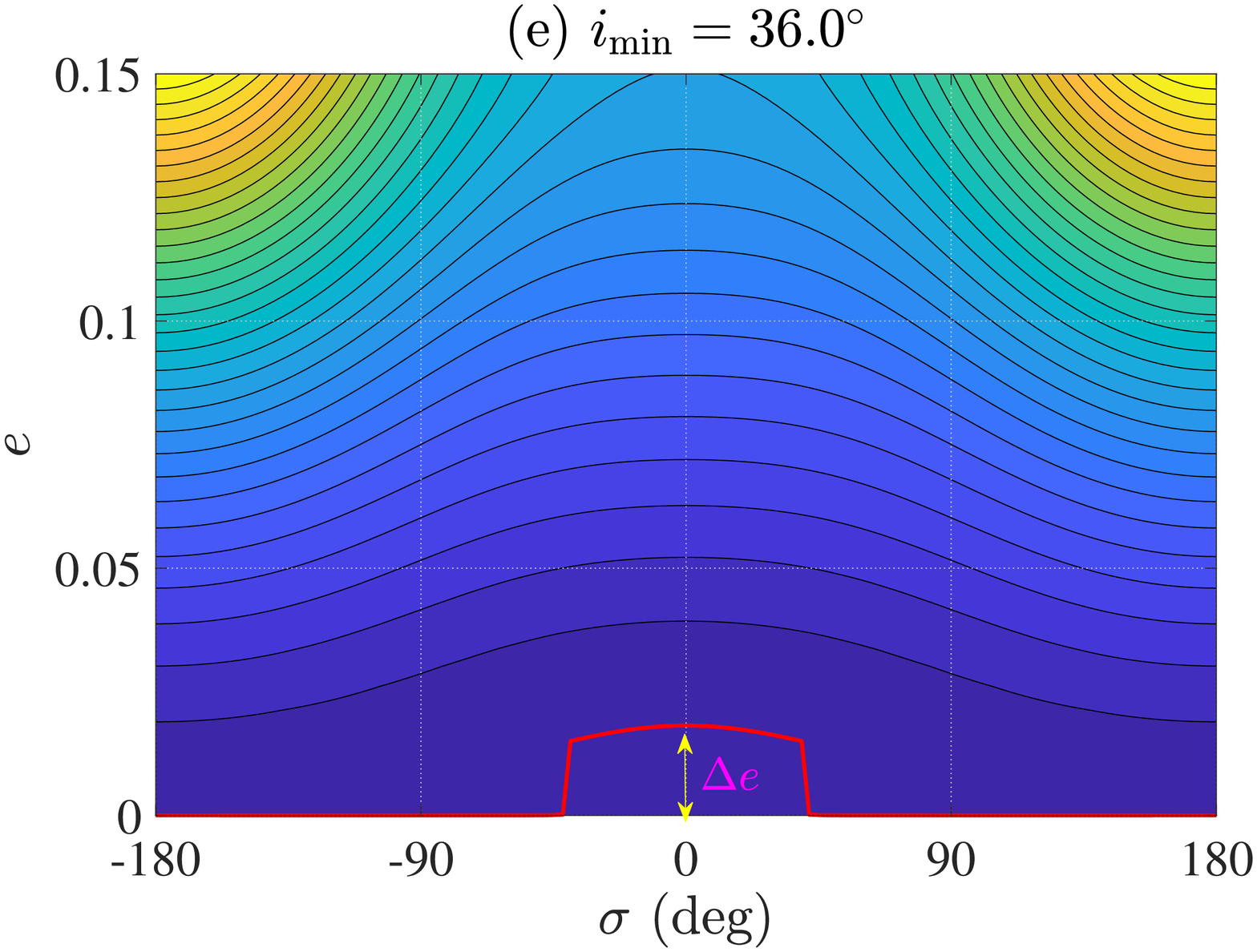}
\includegraphics[width=0.45\textwidth]{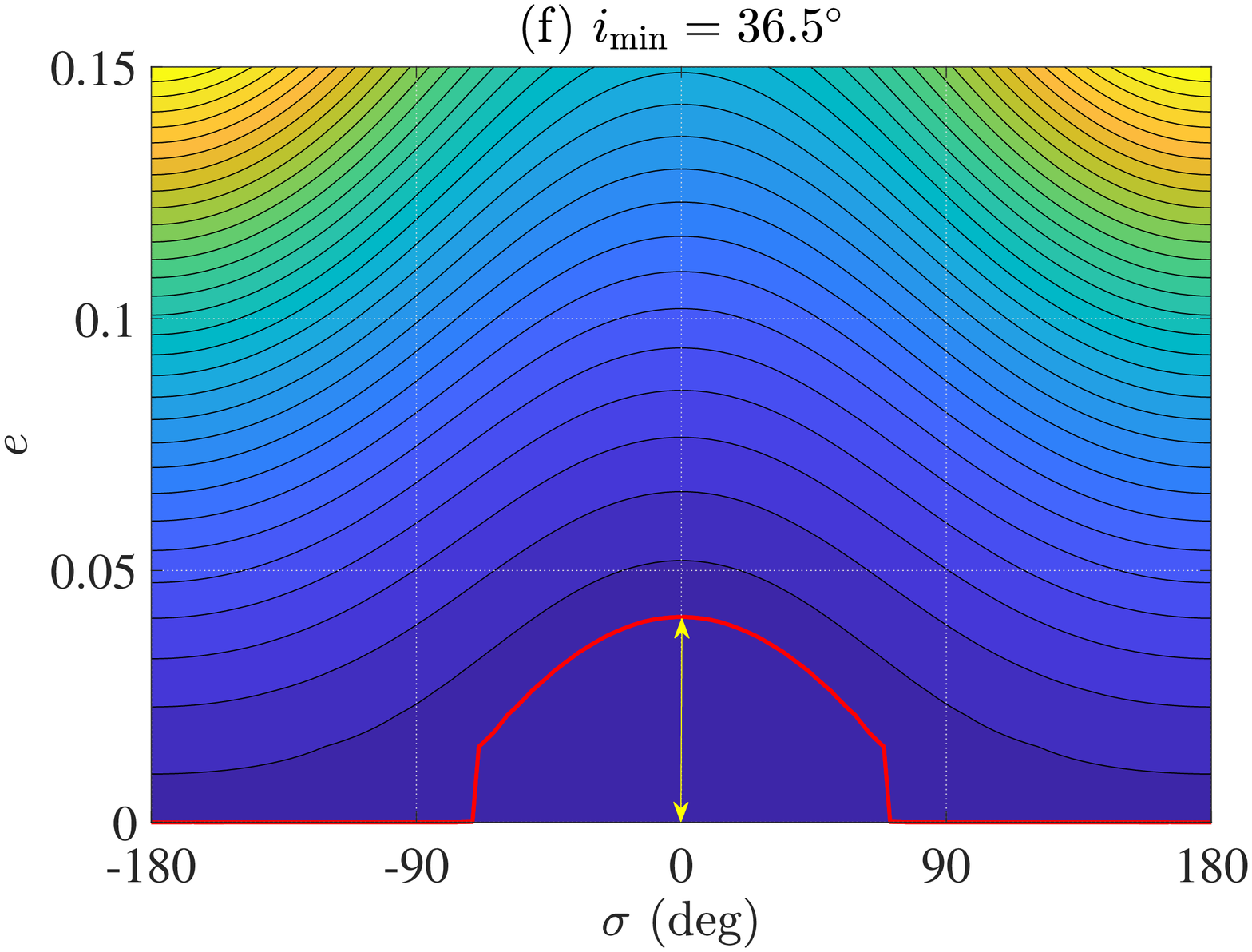}
\caption{Level curves of the resonant Hamiltonian ${\cal H}^*$ for the parameter of minimum inclination $i_{\min}$ near the critical value. In simulations, the semimajor axis ratio is taken as $\alpha = 0.1$. The level curves separating the libration regions from circulation regions are shown in red lines. The size of libration island is measured by the variation of eccentricity, denoted by $\Delta e$.}
\label{Fig8}
\end{figure*}

In the previous subsection, we have found that the curve of resonant width as a function of inclination has an interrupt at the critical inclination $\sim$$35.8^{\circ}$. What happens at such a critical inclination? To make clear this problem, we assume the minim inclination at $35.6^{\circ}$, $35.7^{\circ}$, $35.8^{\circ}$, $35.9^{\circ}$, $36.0^{\circ}$ and $36.5^{\circ}$ and produce the associated phase portraits in the $(\sigma, e)$ space. The phase portraits are reported in Fig. \ref{Fig8}.

When the minimum inclination $i_{\min}$ is at $35.6^{\circ}$, the dynamical structures arising in the phase portrait is similar to the ones shown in Figs. \ref{Fig5} and \ref{Fig6} (there is only one island of resonance in the phase space). Similarly, the resonant width is marked and denoted by $\Delta e$, which can be used to estimate the maximum variation of eccentricity.

When the minimum inclination $i_{\min}$ increases up to $35.7^{\circ}$, besides the primary island of libration entered at $\sigma = 0^{\circ}$, two small islands of libration bifurcate and their centres of resonance are at $\sigma = \pm 180^{\circ}$. In the case of $i_{\min} = 35.7^{\circ}$, the dynamics of test particles initially placed on quasi-circular orbits are dominated by the primary island of libration, so that the maximum variation of eccentricity in this region can be estimated by the size of the primary island of libration.

However, when the minimum inclination $i_{\min}$ is up to $35.8^{\circ}$ (close to the critical inclination), the dynamical structures in the phase portrait change in terms of the following two aspects: (a) the primary island of libration centred at $\sigma = 0^{\circ}$ moves upward and this island of libration cannot dominate the dynamics of particles initially placed on quasi-circular orbits, (b) in the low-eccentricity region there appears a new island of libration centred at $\sigma = \pm 180^{\circ}$. Furthermore, when the minimum inclination $i_{\min}$ is increased up to $35.9^{\circ}$, the primary island of libration disappears and in the low-eccentricity region there is a new island of libration. We can see that, in both the cases of $i_{\min} = 35.8^{\circ}$ and $i_{\min} = 35.9^{\circ}$, the dynamics of particles on quasi-circular orbits are dominated by the islands of libration arising in the low-eccentricity region. However, the size of the libration island arising in the low-eccentricity region is very small. Thus, the maximum variation of eccentricity at $i_{\min} = 35.8^{\circ}$ and $i_{\min} = 35.9^{\circ}$ should be very small due to the small size of the libration islands.

When the minimum inclination $i_{\min}$ grows up to $36.0^{\circ}$ and $36.5^{\circ}$, the size of the libration islands in the low-eccentricity region increases, showing that the maximum variation of eccentricity should be increased.

Based on the aforementioned discussions, we can conclude that the interrupt of $\Delta e$ occurring at the critical inclination $\sim$$35.8^{\circ}$ is due to the change of dynamical structures in phase portraits. In particular, when the minimum inclination $i_{\min}$ becomes greater than the critical inclination, new island of libration appears in the low-eccentricity region and it begins to dominate the the dynamics of particles initially placed on quasi-circular orbits.

\section{Conclusions}
\label{Sect6}

In this work, we investigated the secular resonance associated with the critical argument of $\sigma = \varpi$ for inner test particles in hierarchical planetary systems in order to provide a dynamical explanation for the shape of $\Delta e$ numerically determined in the inclination interval $(0^{\circ},39^{\circ})$ under the full secular model.

The double-averaged Hamiltonian is formulated up to an arbitrary order in the semimajor axis ratio $\alpha$ between the test particle and the disturbing object. Under the full secular model, we reproduced the maximum variation of eccentricity ($\Delta e$) as a function of initial inclination for those test particles initially placed on quasi-circular orbits. We concentrated our attention on the dynamical mechanism hidden in the shape of $\Delta e$.

To understand the secular dynamics, the perturbation theory based on Lie-series transformation is adopted. By taking advantage of the Hori--Deprit method, the `short-period' terms in the double-averaged Hamiltonian are removed, and the resulting resonant model reduces to a single-degree of freedom system with the resonant angle $\sigma$ as the angular coordinate. Based on the resonant Hamiltonian, the phase-space structures, resonant centres, and resonant widths are identified analytically. Results show that both the curves of resonant centre and resonant width have an interrupt at inclination of $\sim$$35.8^{\circ}$, which is due to the change of dynamical structures in phase portraits. In particular, we found a perfect correspondence between the resonant width in terms of the eccentricity variation and the maximum variation of eccentricity $\Delta e$ numerically determined under the full secular model. It is concluded that the secular dynamics of test particles in the interested region is dominated by the secular resonance considered in this work.

\section*{Acknowledgments}
The author thanks Drs Cristian Beaug\'e and Xiyun Hou for helpful discussions and thanks the anonymous reviewer for his/her comments. This work is financially supported by the National Natural Science Foundation of China (Nos. 12073011) and the National Key Research and Development Program of China (No.2019YFA0706601).

\section*{Data availability}
The analysis and codes are available upon request.

\bibliographystyle{mn2e}
\bibliography{mybib}


%
%


\bsp
\label{lastpage}
\end{document}